\documentclass[journal]{IEEEtran}
\usepackage{graphicx}
\usepackage{amsmath}
\usepackage{float}
\usepackage{caption}
\usepackage{subcaption}
\usepackage{dblfloatfix}
\usepackage[colorlinks=true,linktocpage=false,linkcolor=blue,citecolor=blue,urlcolor=black,bookmarksopen=true]{hyperref}
\usepackage{bookmark}
\usepackage[export]{adjustbox}
\usepackage[utf8]{inputenc}

\ifCLASSINFOpdf
\else
\fi
\begin{document}
%
\title{Broadband Multifunctional Plasmonic Polarization Converter based on Multimode Interference Coupler}
%
%
%

\author{Hamed~Pezeshki,
        Bert~Koopmans,
        and~Jos~J.~G.~M.~van~der~Tol
\thanks{This work is part of the Gravitation program ‘Research Centre for Integrated Nanophotonics’, which is financed by the Netherlands Organization for Scientific Research (NWO). (Corresponding author: Hamed Pezeshki.)}
\thanks{H. Pezeshki and Bert Koopmans are with the Department of Applied Physics, Eindhoven University of Technology, Eindhoven, 5612 AZ, Netherlands (email: h.pezeshki@tue.nl).}
\thanks{H. Pezeshki, Bert Koopmans, and Jos J. G. M. van der Tol are with the Eindhoven Hendrik Casimir Institute, Center for Photonic Integration, Eindhoven University of Technology, Eindhoven, 5600 MB, Netherlands.}
}

\maketitle

\begin{abstract}
We propose a multifunctional integrated plasmonic-photonic polarization converter for polarization demultiplexing in an indium-phosphide membrane on silicon platform. Using a compact 1$\times$4 multimode interference coupler, this device can provide simultaneous half-wave plate and quarter-wave plate (HWP and QWP) functionalities, where the latter generates two quasi-circular polarized beams with opposite spins and topological charges of $l$ = $\pm$1. Our device employs a two-section HWP to obtain a very large conversion efficiency of $\geq$ 91\% over the entire C to U telecom bands, while it offers a conversion efficiency of $\geq$ 95\% over $\sim$ 86\% of the C to U bands. Our device also illustrates QWP functionality, where the transmission contrast between the transverse electric and transverse magnetic modes is $\approx$ 0 dB over the whole C band and 55\% of the C to U bands. We expect this device can be a promising building block for the realization of ultracompact on-chip polarization demultiplexing and lab-on-a-chip biosensing platforms. Finally, our proposed device allows the use of the polarization and angular momentum degrees of freedom, which makes it attractive for quantum information processing.
\end{abstract}

\begin{IEEEkeywords}
Plasmonics, Polarization converter, half-wave plate, quarter-wave plate, Photonic integrated circuit, Indium phosphide.
\end{IEEEkeywords}

%
\IEEEpeerreviewmaketitle

\IEEEPARstart{I}{ndium-phosphide} (InP) membrane on silicon (IMOS) is a promising platform for the fabrication of low cost and large scale passive and active photonic integrated circuits (PICs) \cite{van2017indium,van2019inp}, due to its compatibility with complementary metal oxide semiconductor (CMOS) processes. The high refractive index difference (n $\approx$ 2) between its core and cladding enables large scale integration of photonic devices \cite{kish2011current} and photonics-electronics convergence on a single chip \cite{jiao2020inp}. However, the large refractive index difference makes it harder to obtain polarization independence in photonic devices \cite{dai2013polarization}. Meanwhile, control over the polarization in PICs for polarization-independent operations of chips and for functions like polarization (de)multiplexing and polarization switching, is of great importance \cite{bogaerts2007polarization}. Hence, the development of an integrated polarization converter (PC) with a small footprint to support polarization diversity, as well as to convert the polarization state of light in PICs at will, have attracted a lot of attention during recent decades.\par

So far, several design proposals have been made, based on either mode evolution or mode interference. PCs based on mode evolution implement adiabatic mode conversion between the two polarization states of a waveguide mode \cite{wang2014novel,keyvaninia2019highly}, and are typically long ($\geq$ 100 $\mu$m). In contrast, shorter PCs use the interference between two orthogonal beating modes, propagating through a symmetry-broken waveguide \cite{zhang2011ultracompact,caspers2012compact}. The majority of the proposed PCs based on mode interference were developed with slanted waveguides \cite{jiang2018fabrication} and narrow trenches \cite{nakayama2012single}. However, such approaches entail either a large footprint or a complex fabrication process \cite{gao2013ultra}.\par

It has been demonstrated that a plasmonic metal layer can overcome the above-mentioned challenges by enhancing the birefringence between the two beating modes using surface plasmon polaritons (SPPs), as well as offering a simple fabrication process. Early works on designs of integrated half-wave plates (HWPs) based on metal layers, show that large ohmic losses \cite{zhang2011ultracompact}, caused by SPPs, can be decreased by placing a low-refractive index thin spacer layer at the metal-dielectric interface \cite{caspers2012compact,komatsu2012compact}. Komatsu \textit{et al.} \cite{komatsu2012compact} and Caspers \textit{et al.} \cite{caspers2012compact} presented HWPs with insertion losses (ILs) of $\sim$5 and $>$ 2 dB for the device lengths of 11 and 5 $\mu$m, respectively. Later on, other groups reported SPP-based HWPs with a high polarization conversion efficiency (PCE) of 97\% with an IL of $\sim$2 dB. Despite works done so far, broadband and multifunctional operation has not been addressed sufficiently yet.\par 

There have also been some reports on the integrated quarter-wave plates (QWPs) using a plasmonic metal layer. Gao \textit{et al.} \cite{gao2015chip} theoretically presented an integrated hybrid QWP based on plasmonics at $\lambda_0$=1.55 $\mu$m, where the PC's length was 1.5 $\mu$m. Liang \textit{et al.} \cite{liang2016integratable} theoretically showed a QWP with one-way angular momentum conversion at $\lambda_0$=1.55 $\mu$m, by placing a L-shaped metal layer with a length of 2.8 $\mu$m on a square photonic waveguide with a minimum birefringence, which is attached to a 2.4 $\mu$m long rectangular photonic waveguide with high birefringence. However, as shown in this paper, a better strategy would be to place the metal layer on a rectangular birefringence waveguide to further boost the birefringence for a shorter length of the metal layer. This, in turn, results in lower absorption loss (i.e. heat dissipation) by the metal layer, which is crucial for the performance efficiency of devices on a photonic chip. Moreover, there have been some reports on the design of an integrated QWP based on aluminum gallium arsenide (AlGaAs) with an ellipticity of 0.67 at $\lambda_0$=1.55 $\mu$m \cite{maltese2018towards} as well as with graphene in Terahertz regime \cite{ni2019selective} by launching linearly polarized light at 45$^\circ$. Both designs have very long converter sections of $\sim$53 $\mu$m and 145 $\mu$m, respectively. However, in the latter case, they achieved active adjustment of the polarization state of light through variation of the graphene's Fermi level. Despite several works done so far on both integrated HWP and QWP, design of a broadband multifunctional PC, to provide both polarization and angular momentum degrees of freedom has not been investigated.\par

In this paper, we introduce a multifunctional PC with an ultrabroad operational wavelength range. After first designing a compact and efficient 1$\times$4 multimode interference coupler (MMI), our proposed HWP is designed as a two-section PC to achieve optimum conversion efficiency and to improve fabrication tolerance as demonstrated in \cite{van2012increasing}. According to results obtained with a finite-difference time-domain (FDTD) method \cite{fdtd}, the proposed HWP offers a PCE of $\geq$ 91\% over the C to U telecom bands, while PCE is $\geq$ 95\% over $\sim$ 86\% of this wavelength range. This implies that our HWP presents a polarization extinction ratio (PER) of better than 13 dB in the above-mentioned range, with a maximum of 38.4 dB at $\lambda_0$= 1.563 $\mu$m.

By taking advantage of the mirror symmetry in a MMI, we then propose two QWPs, which are mirrored to each other, presenting quasi-circular polarized beams with opposite spins (due to the transverse spin angular momentum, SAM) based on only one device, one input polarization, and one incoming light beam direction. Our QWPs function efficiently by offering a transmission contrast of $\approx$ 0 dB between the transverse electric and transverse magnetic (TE\textsubscript{0} and TM\textsubscript{0}) modes over the wavelength range of $\lambda_0$=1.53 to 1.61 $\mu$m, covering the whole C band and moreover 55\% of the C to U telecom bands. The longitudinal electric field component of the generated quasi-circular polarized beams carry longitudinal orbital angular momentum (OAM) with topological charges of $l$ = $\pm$1. Having two quasi-circular polarized beams simultaneously on a chip can be potentially attractive not only for on-chip telecom applications such as mode/polarization-(de)multiplexing \cite{pezeshki2022highly}, on-chip magneto-plasmonics \cite{pezeshki2022optical,pezeshki2022design}, as well as quantum information processing \cite{flamini2018photonic}, but also for biosensing applications including circular dichroism spectroscopy \cite{wang2017circular} and nanoparticle movement \cite{kawata1996optically} using a longitudinal OAM.\par

\section{Design Structure and Considerations} \label{s-Design}

\begin{figure}[h!]
             \centering
                \begin{subfigure}[t!]{0.5\textwidth}
                \centering
                 \includegraphics[scale=0.3]{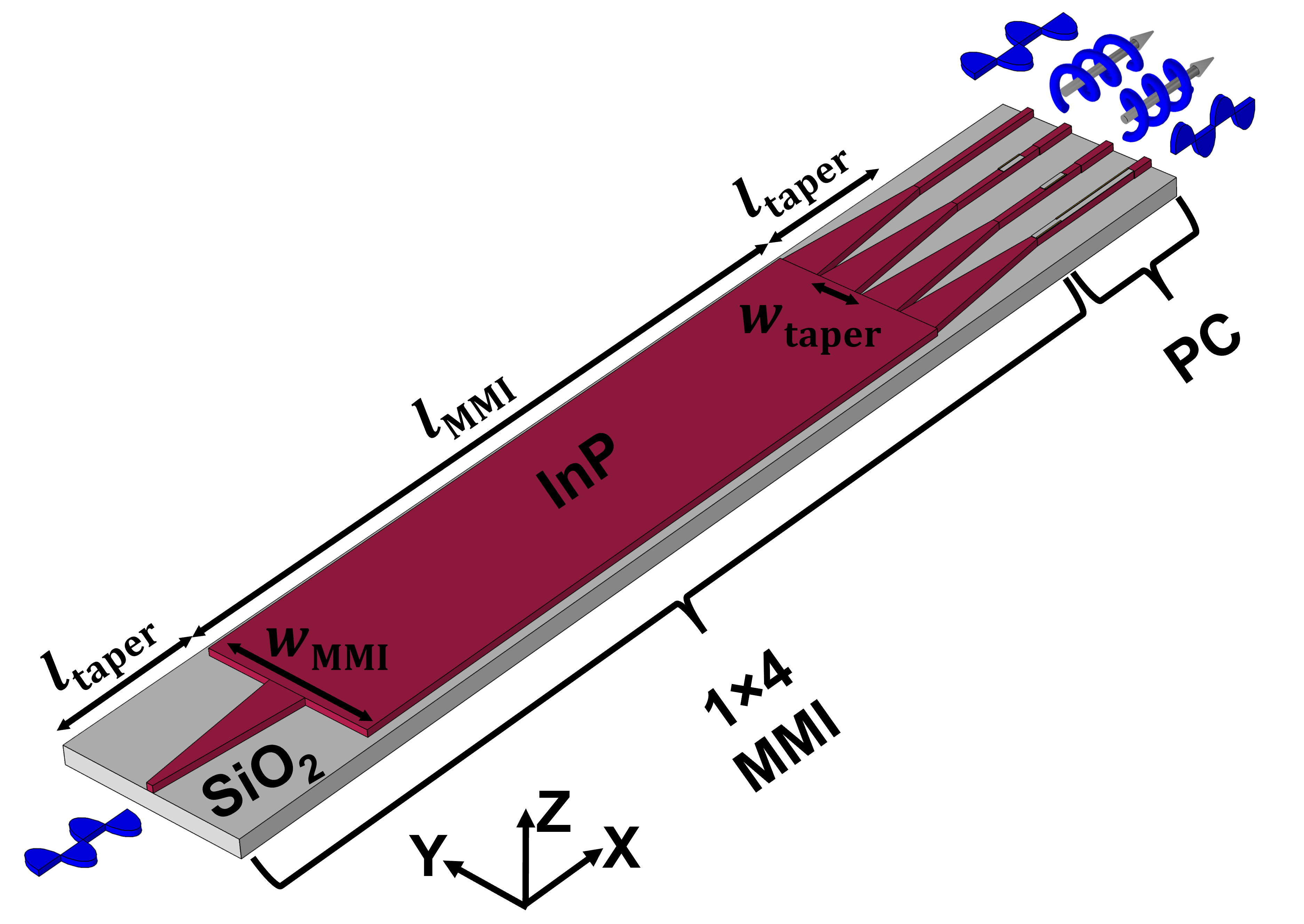}
                 \caption{\vspace{0.1cm}}
                   \label{fig 1a}
                 \end{subfigure}
                 \begin{subfigure}[t!]{0.49\columnwidth}
                \centering
                 \includegraphics[scale=0.18,left]{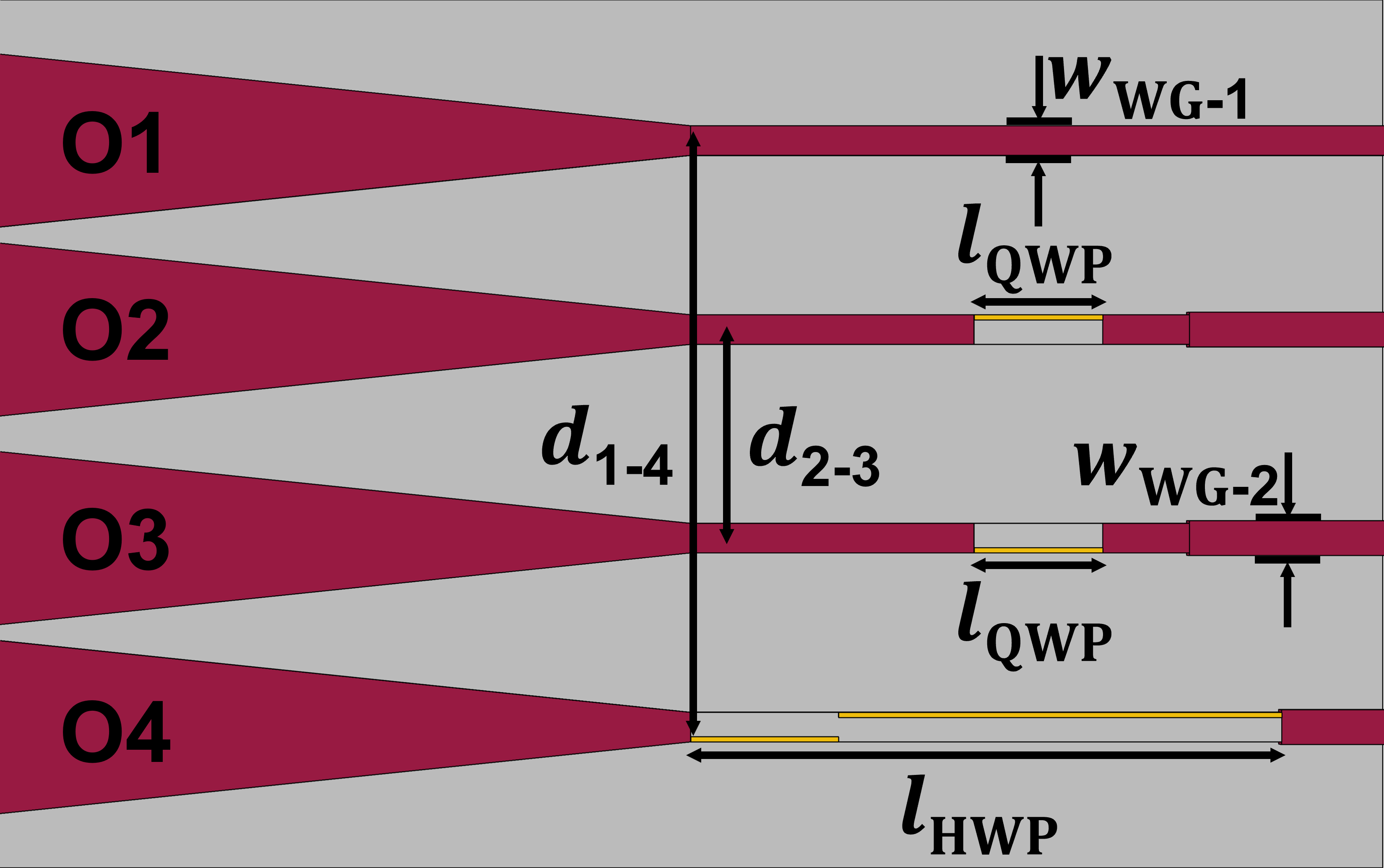}
                \caption{}
                 \label{fig 1b}
                \end{subfigure}
                \begin{subfigure}[t!]{0.49\columnwidth}
                \centering
                 \includegraphics[scale=0.15,right]{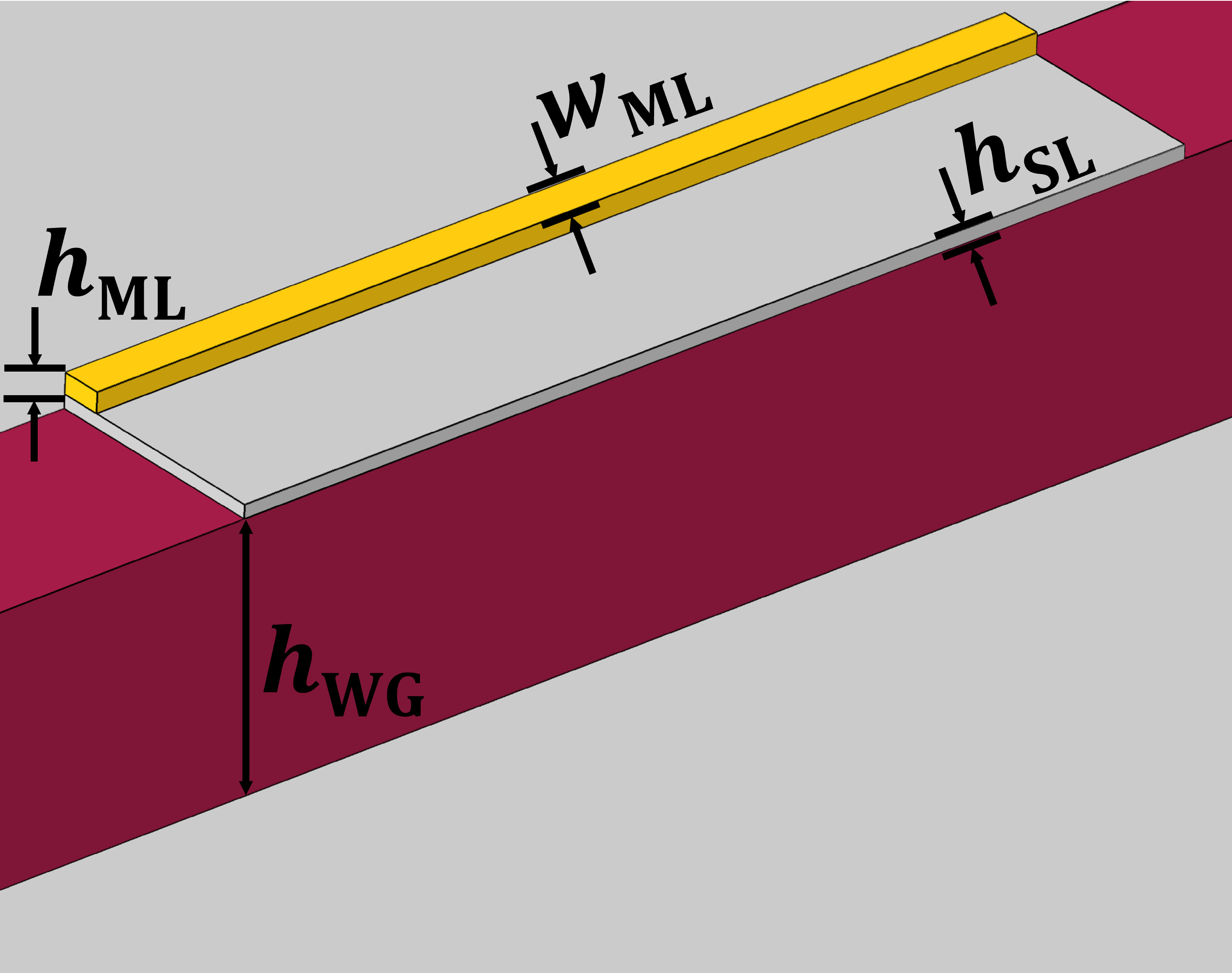}
                  \caption{\hspace*{-0.5cm}}
                  \label{fig 1c}
                 \end{subfigure}
                  \centering
                \caption{Schematic of the proposed multifunctional polarization converter (PC). (a) Perspective view that shows the device is composed of a 1 $\times$ 4 multimode interference coupler (MMI) and a PC section. The lengths and widths of the MMI and tapers are indicated with $l$\textsubscript{MMI}, $l$\textsubscript{taper}, $w$\textsubscript{MMI}, and $w$\textsubscript{taper}, respectively. (b) Top view of the PC section showing the distances between the output waveguides O1(2) and O4(3) by $d_{1-4}$ and $d_{2-3}$, respectively. The widths of the waveguide sections are presented by $w$\textsubscript{WG-1} and $w$\textsubscript{WG-2} and the lengths of the quarter-wave plates (QWPs) and half-wave plate (HWP) are shown by $l$\textsubscript{QWP} and $l$\textsubscript{HWP}, respectively. (c) Perspective view of a PC in which $w$\textsubscript{ML} and $h$\textsubscript{ML} are the width and height of the metal layer of the PC, $h$\textsubscript{SL} and $h$\textsubscript{WG} are the heights of the spacer layer and waveguide, respectively.}
                \label{fig 1}
                \end{figure}

We propose a device based on the IMOS platform to provide different polarizations: linear TE\textsubscript{0} and TM\textsubscript{0} modes, as well as two quasi-circular polarized beams with opposite spins. It is composed of a 1$\times$4 MMI for dividing a TE\textsubscript{0}-polarized light into four output waveguides, as well as a PC section for performing HWP and QWP functionalities (see Fig. \ref{fig 1}). The output waveguide 1 (O1) is connected to a rectangular waveguide and outputs a TE\textsubscript{0} mode. To obtain the quasi-circular polarized beams with opposite spins, we designed two plasmonic QWPs on top of the O2 and O3 waveguides, which are mirrored relative to each other. These created states will exist over propagation lengths that are much smaller than the TE\textsubscript{0}-TM\textsubscript{0} beat length of $\sim$ 77.5 $\mu$m for a $\Delta n$\textsubscript{TE\textsubscript{0}-TM\textsubscript{0}} $\sim$ 0.02. The advantage of our approach over the previous approaches is that we illustrate QWPs with opposite spins using one device, without reversing the direction of light propagation. This makes it attractive for applications that require both spins simultaneously on a single chip. Finally, to obtain a TM\textsubscript{0} mode, we designed a plasmonic HWP, as a top cladding on the O4 waveguide. The proposed HWP is a two-section PC which consists of a combination of a QWP and a three quarter-wave plate (TQWP). We chose this approach to achieve our objectives of highly efficient conversion efficiency over an ultrabroad wavelength range, as well as enhanced fabrication tolerance, as demonstrated in \cite{van2012increasing}. Using this approach, the length of the converter is two times that of a single-section HWP. \par

The MMI section has a length and a width of $l$\textsubscript{MMI} = 36 $\mu$m and $w$\textsubscript{MMI} = 9 $\mu$m, respectively. In order to have an ultralow loss 1$\times$4 MMI, we provide the coupling between the MMI section and the input as well as the output waveguides using linear tapers with a length and a width of $l$\textsubscript{taper} = 8 $\mu$m and $w$\textsubscript{taper} = 2 $\mu$m, respectively, creating a smooth mode transitions between the input, MMI, and output sections. Fig. \ref{fig 1b} shows the top view of the PC section in which the center-to-center distances between the O1(2) and O4(3) waveguides are $d_{1-4}$ = 6.75 and $d_{2-3}$ = 2.4 $\mu$m, respectively. The QWPs devices on the O2 and O3 waveguides have a plasmonic cladding with a length of $l$\textsubscript{QWP} = 1.48 $\mu$m to transform a TE\textsubscript{0} mode into quasi-circular polarized beams with two opposite spins. The designed HWP, with a total length of $l$\textsubscript{HWP} = 6.8 $\mu$m, is designed as a top cladding on the O4 waveguide to transform a TE\textsubscript{0} mode to a TM\textsubscript{0} mode. The input and O1 waveguides have a width of $w$\textsubscript{WG-1} = 0.34 $\mu$m, while the O2 to O4 waveguides are linked with waveguides with a different width of $w$\textsubscript{WG-2} = 0.4 $\mu$m to minimize the birefringence in end parts of the waveguides, i.e. $\Delta n$\textsubscript{TE\textsubscript{0}-TM\textsubscript{0}} $\sim$ 0.02. The height of all photonic components are equal to $h$\textsubscript{WG} = 0.39 $\mu$m. The width and height of all the plasmonic metal layers, as well as the height of silica (SiO\textsubscript{2}) spacer layer are the same in the O2 to O4 waveguides, i.e. $w$\textsubscript{ML} = 0.03 $\mu$m, $h$\textsubscript{ML} = 0.03 $\mu$m, and $h$\textsubscript{SL} = 0.02 $\mu$m, respectively (see Fig. \ref{fig 1c}). As indicated in Fig. \ref{fig 1}, the materials for the waveguides and MMI, substrate and the spacer layer beneath the plasmonic components are InP and SiO\textsubscript{2}, whose parameters come from \cite{palik1998handbook}, and the material for the plasmonic components is gold \cite{johnson1972optical}.\par

We evaluate the performance of the proposed HWP in section \ref{ss-HWP} based on PCE ($\eta$ in percent) and insertion loss (\textit{IL}\textsubscript{H} in dB). For an input TE\textsubscript{0} mode, we have:

            \begin{equation}
                \eta = ({P^{~\mathrm{O4}}_{\mathrm{TM}}} / ({P^{~\mathrm{O4}}_{\mathrm{TM}}} + {P^{~\mathrm{O4}}_{\mathrm{TE}}})) \times 100,
                \label{Eq-1}
            \end{equation}
            
            \begin{equation}
                \mathit{IL}_\mathrm{H} = -10 \times \log (P^{~\mathrm{O4}}_{\mathrm{TM}}/P^{~\mathrm{I4}}_{\mathrm{TE}}),
                \label{Eq-3}
            \end{equation}
            
\noindent where $P^{~\mathrm{I4}}_{\mathrm{TE}}$ and $P^{~\mathrm{O4}}_{\mathrm{TE}}$ are the TE\textsubscript{0} mode powers at the input and output of the O4 waveguide, while $P^{~\mathrm{O4}}_{\mathrm{TM}}$ is the TM\textsubscript{0} mode power at the output of the O4 waveguide. Note that in calculating IL using Eq. \ref{Eq-3}, the splitting loss by the MMI is neglected.\par

In section \ref{ss-QWP}, we assess the presented QWP function according to the transmission contrast, TC, (\textit{C}\textsubscript{T} in dB) between the beating TE\textsubscript{0} and TM\textsubscript{0} modes and insertion loss (\textit{IL}\textsubscript{Q} in dB). Since both QWPs are identical and just mirrored relative to each other, we will just illustrate the results for the O2 waveguide, where \textit{C}\textsubscript{T} and \textit{IL}\textsubscript{Q} are defined as:

            \begin{equation}
                C_T = 10 \times \log (T^{~\mathrm{O2}}_{\mathrm{TM}} /T^{~\mathrm{O2}}_{\mathrm{TE}}),
                \label{Eq-4}
            \end{equation}

            \begin{equation}
                \mathit{IL}_\mathrm{Q} = -10 \times \log ((P^{~\mathrm{O2}}_{\mathrm{TE}} + P^{~\mathrm{O2}}_{\mathrm{TM}}) /P^{~\mathrm{I2}}_{\mathrm{TE}}),
                \label{Eq-5}
            \end{equation}

\noindent where $T^{~\mathrm{O2}}_{\mathrm{TM}}$ and $T^{~\mathrm{O2}}_{\mathrm{TE}}$ are the transmissions of TM\textsubscript{0} and TE\textsubscript{0} modes through the O2 waveguide, $P^{~\mathrm{O2}}_{\mathrm{TE}}$ and $P^{~\mathrm{O2}}_{\mathrm{TM}}$ are the TE\textsubscript{0} and TM\textsubscript{0} mode powers at the output of the O2 waveguide, while $P^{~\mathrm{I2}}_{\mathrm{TE}}$ is the TE\textsubscript{0} mode power at the input of the O2 waveguide. In calculating IL using Eq. \ref{Eq-5}, the splitting loss by the MMI is also neglected.\par

\section{Numerical Results} \label{s-results}

\subsection{MMI Performance} \label{ss-MMI}

\begin{figure}[h!]
             \centering
                \begin{subfigure}[t!]{0.5\textwidth}
                \centering
                 \includegraphics[scale=0.3]{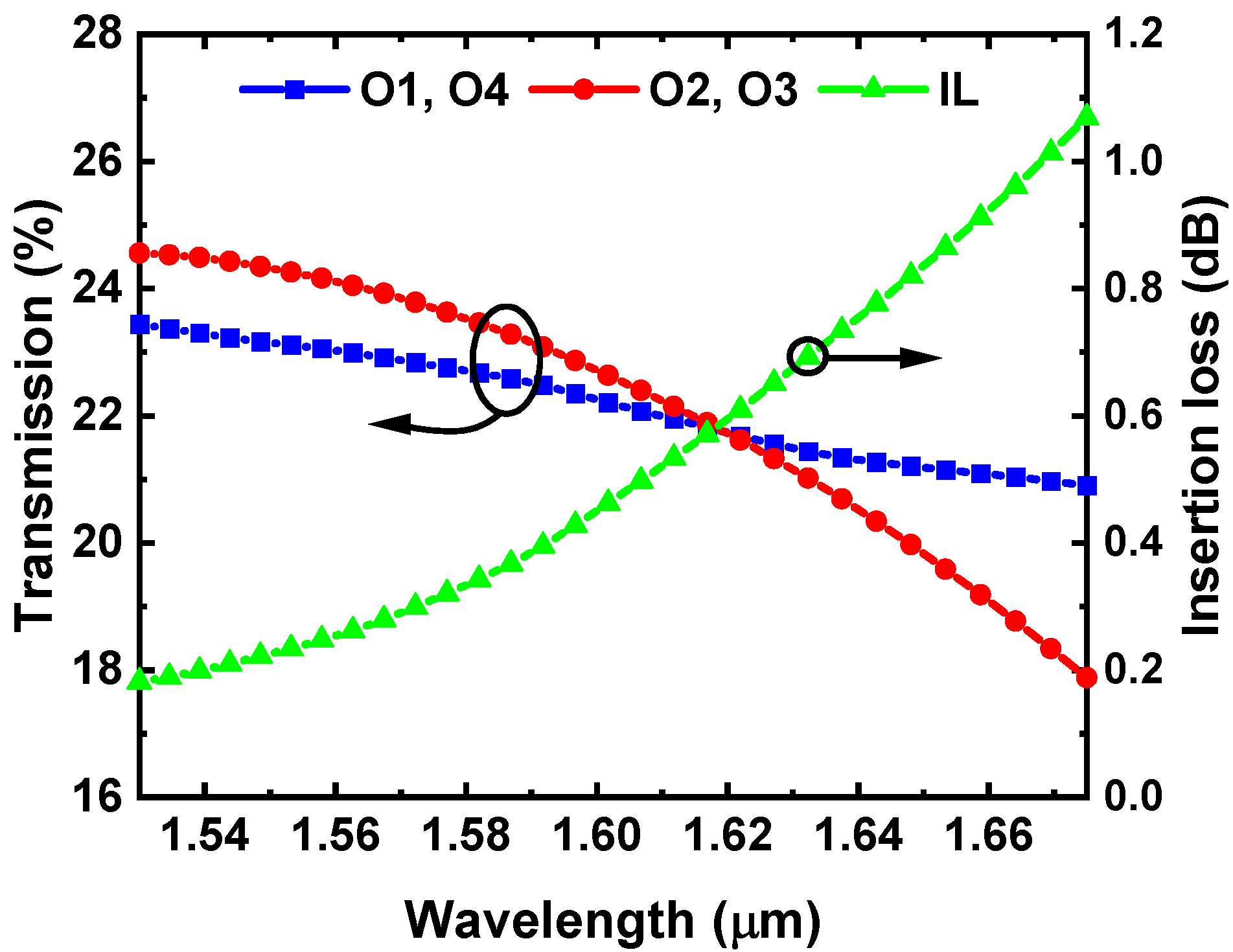}
                 \caption{}
                   \label{fig 2a}
                 \end{subfigure}
                 \begin{subfigure}[t!]{0.49\columnwidth}
                \centering
                 \includegraphics[scale=0.26,right]{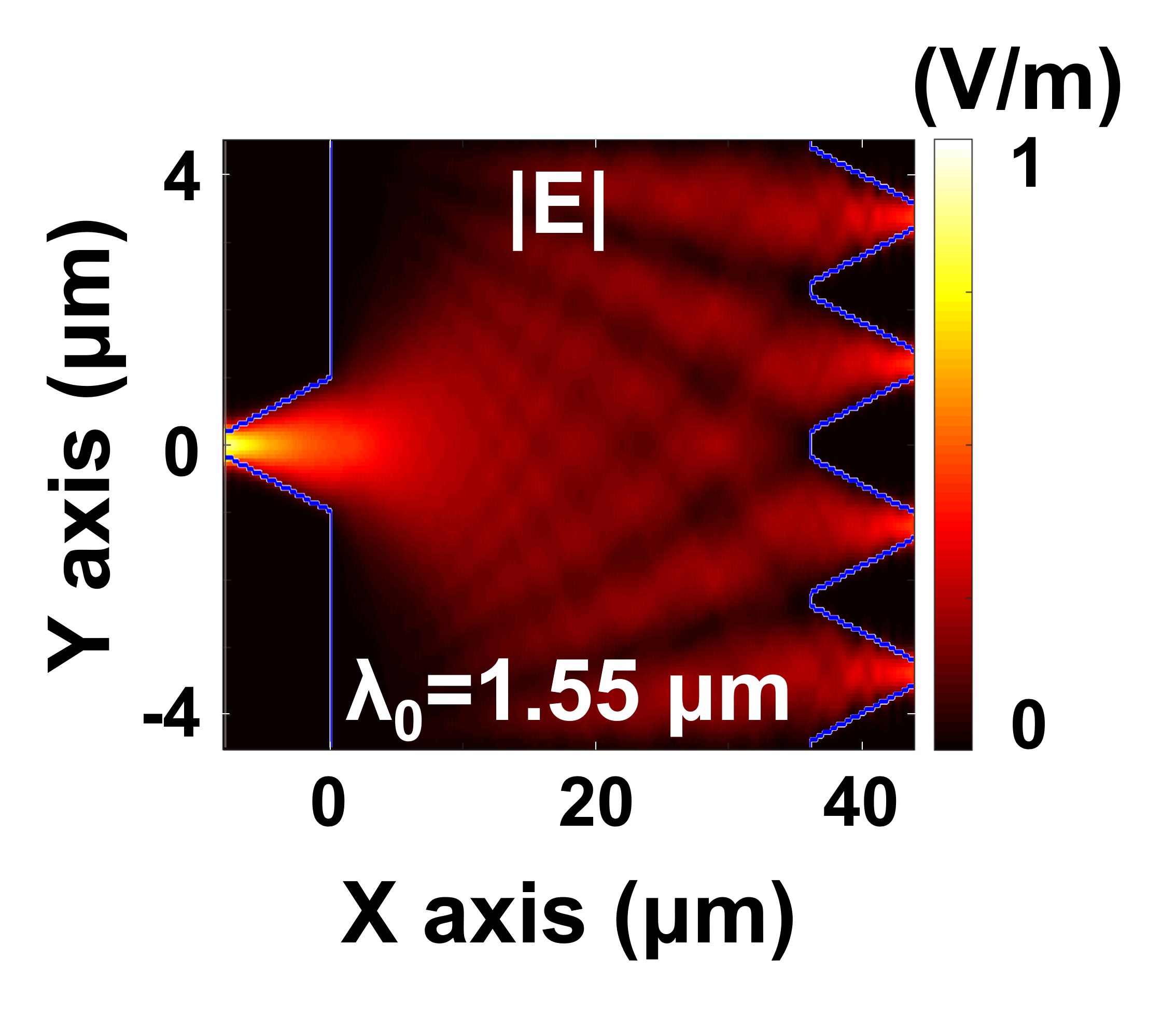}
                \caption{}
                 \label{fig 2b}
                \end{subfigure}
                \begin{subfigure}[t!]{0.49\columnwidth}
                \centering
                 \includegraphics[scale=0.26,left]{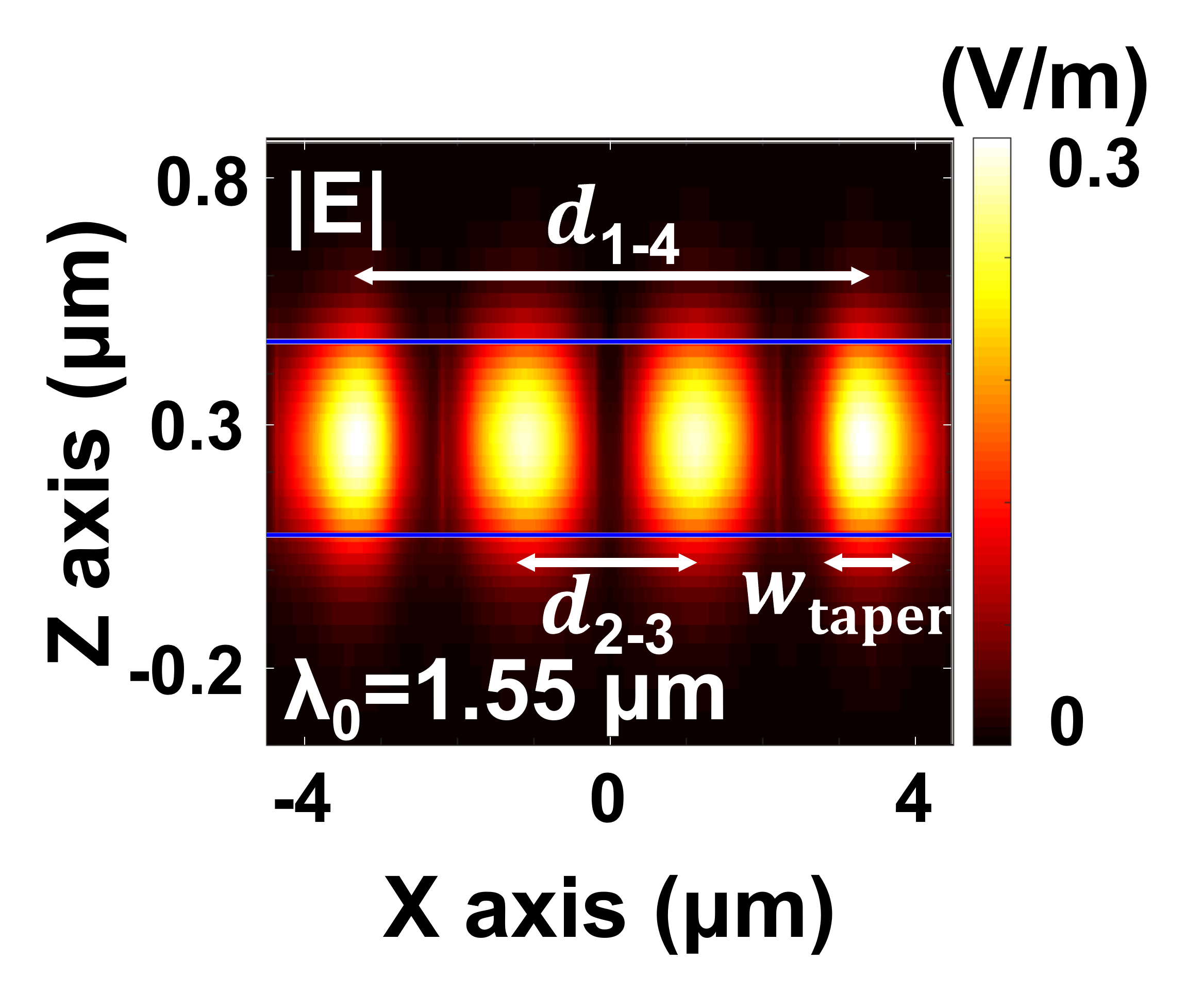}
                \caption{}
                 \label{fig 2c}
                \end{subfigure}
                \centering
                \caption{Performance of the MMI section over the telecom wavelength range of C to U, i.e. $\lambda_0$ = 1.53 to 1.675 $\mu$m. (a) Transmission of individual output waveguides, O1 to O4, and the insertion loss (IL) of MMI at the left and right \textbf{Y} axes, respectively. (b, c) The two dimensional (2D) electric field distribution across the MMI section in the \textbf{XY} plane, and at the end of MMI in the \textbf{YZ} plane. $w$\textsubscript{taper}, $d_{1-4}$, and $d_{2-3}$ are the width of the taper and distances between the outputs 1-4 and 2-3, respectively.}
                \label{fig 2}
                \end{figure}

Figure \ref{fig 2a} shows the performance of the MMI over C to U telecom bands. According to this figure, all outputs have a transmission of $\geq$ 22$\%$ over the wavelength range of 1.53 to 1.62 $\mu$m and the insertion loss is $<$ 0.6 dB. Note that this insertion loss is calculated as the ratio of the sum of the output powers of all outputs to the input power. For the wavelength range of 1.62 to 1.675 $\mu$m, the O1 and O4 waveguides keep a transmission of 21$\%$, while transmission in the O2 and O3 waveguides reduces from 22$\%$ to 18$\%$, which results in an increased insertion loss of $<$ 1.1 dB, which is acceptable. As the MMI section was optimized for an initial wavelength of 1.55 $\mu$m, in Figs. \ref{fig 2b} and \ref{fig 2c} we depicted the two dimensional (2D) electric field distributions in the MMI section at that wavelength. Figure \ref{fig 2b} shows how a single TE\textsubscript{0} mode is split inside the MMI region and converges to four modes at the end of this area. The distribution of the electric field in the \textbf{YZ} plane is shown in Fig. \ref{fig 2c} from which the taper's width and the distance between the outputs, i.e. $w$\textsubscript{taper}, $d_{1-4}$, and $d_{2-3}$, are derived in order to lower the insertion loss as much as possible.\par

\subsection{HWP Functionality} \label{ss-HWP}

            \begin{figure}[ht!]
             \centering
                \includegraphics[scale=0.3]{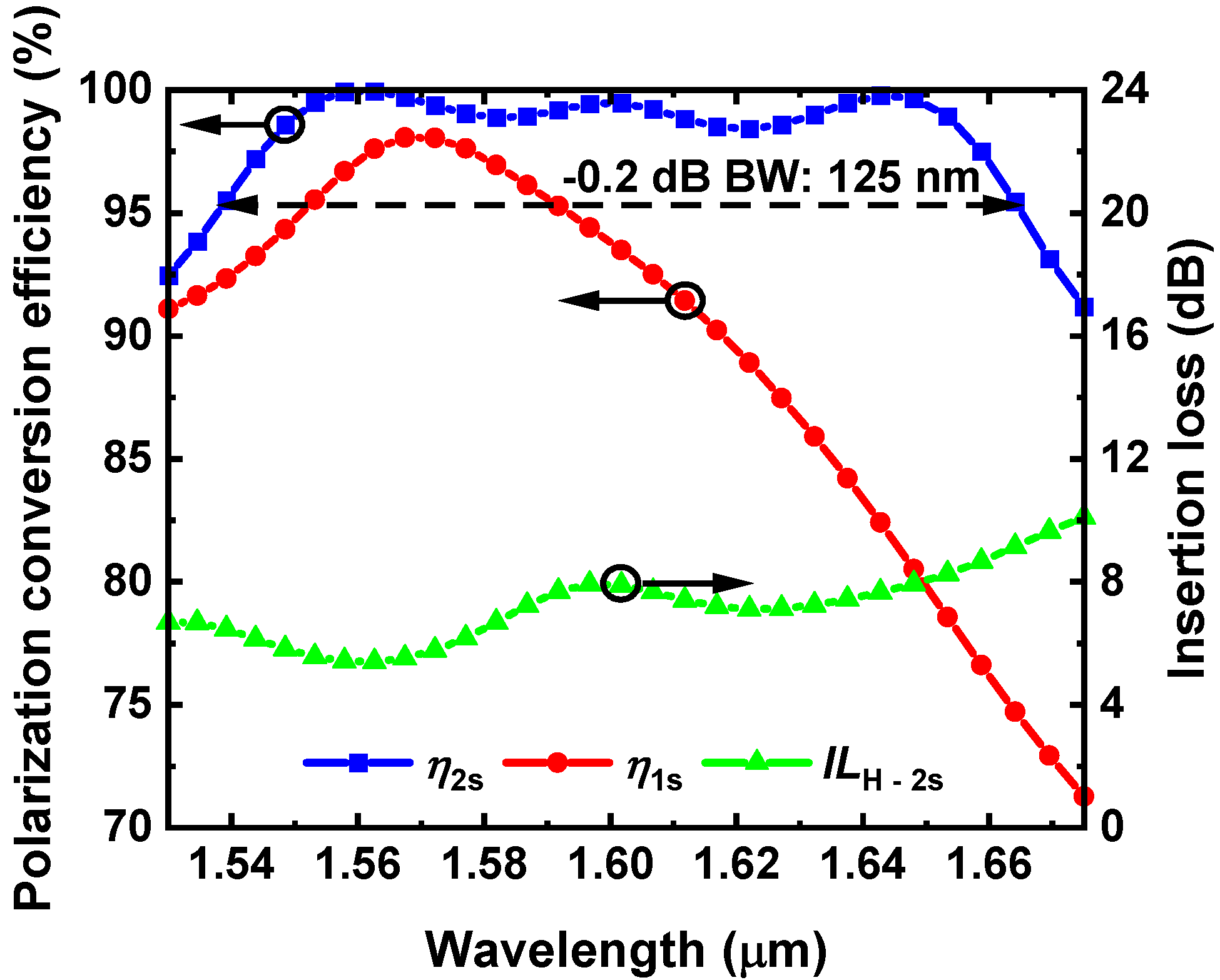}
                \centering
                \caption{Performance of the two-section HWP (denoted by 2s) over the telecom wavelength range of C to U, i.e. $\lambda_0$ = 1.53 to 1.675 $\mu$m. Polarization conversion efficiency and IL ($\eta$\textsubscript{2s} and \textit{IL}\textsubscript{H - 2s}) at the left and right \textbf{Y} axes, respectively. For the comparison, $\eta$\textsubscript{1s} of the one-section HWP is also plotted.}
                \label{fig 3}
                \end{figure}

Figure \ref{fig 3} shows the performance of the two-section HWP over the telecom wavelength range of 1.53 to 1.675 $\mu$m, i.e. C to U bands. Figure \ref{fig 3} shows PCE for both the two-section HWP (denoted by $\eta$\textsubscript{2s}) and the conventional one-section HWP (denoted by $\eta$\textsubscript{1s}) to show the superior performance of our design to the conventional one. As shown in Fig. \ref{fig 3}, $\eta$\textsubscript{2s} is $\geq$ 91\% over the whole C to U bands with a maximum of 99.95\% at $\lambda_0$ = 1.563 $\mu$m. More importantly, the proposed two-section HWP illustrates a 0.2 dB bandwidth of 125 nm for $\eta$\textsubscript{2s} of $\geq$ 95\%, i.e. $\sim$ 86\% of the C to U telecom bands. In contrast, the one-section HWP has a $\eta$\textsubscript{1s} of $\geq$ 91\% for the wavelength range of $\lambda_0$ = 1.53 to 1.615 $\mu$m, even not covering the whole C and L bands. Moreover, this coupler has a 0.2 dB bandwidth of only 39 nm for $\eta$\textsubscript{1s} of $\geq$ 95\%. The comparison of the performance of the conventional HWP with our proposed two-section HWP shows indeed the latter case outperforms the one-section HWP. The very large PCE over 125 nm of the C to U bands makes our proposed two-section HWP device potentially attractive for mode/polarization (de)multiplexing. As the two-section HWP shows its advantage over the conventional one, we continue with the two-section HWP. According to the right \textbf{Y} axis in this figure, IL of our HWP (\textit{IL}\textsubscript{H - 2s}) varies between 5.4 to 7.9 dB with its minimum at $\lambda_0$ = 1.563 $\mu$m. At first glance, the IL range might seem a bit high compared to some recent works on plasmonic HWPs, but we should notice that we used a two-section HWP in contrast to earlier reports. Due to the use of two-section HWP, the length of gold metal layer is $l$\textsubscript{HWP} = 6.8 $\mu$m, which consequently increases the loss by only 2.7 to 4 dB. However, according to the results in Fig. \ref{fig 3} and as demonstrated in \cite{van2012increasing}, from the application perspective of our proposed device, in the trade-off between IL, ultrabroad operation range, much higher PCE, we prioritized the latter two performance parameters. Finally, the reflection from the HWP back to the MMI is calculated, which is lower than -18.8 dB throughout the whole wavelength range, with its minimum value of -29.5 dB at $\lambda_0$ = 1.563 $\mu$m. \par

            \begin{figure}[b!]
             \centering
                \begin{subfigure}[t!]{0.49\columnwidth}
                \centering
                 \includegraphics[scale=0.25,right]{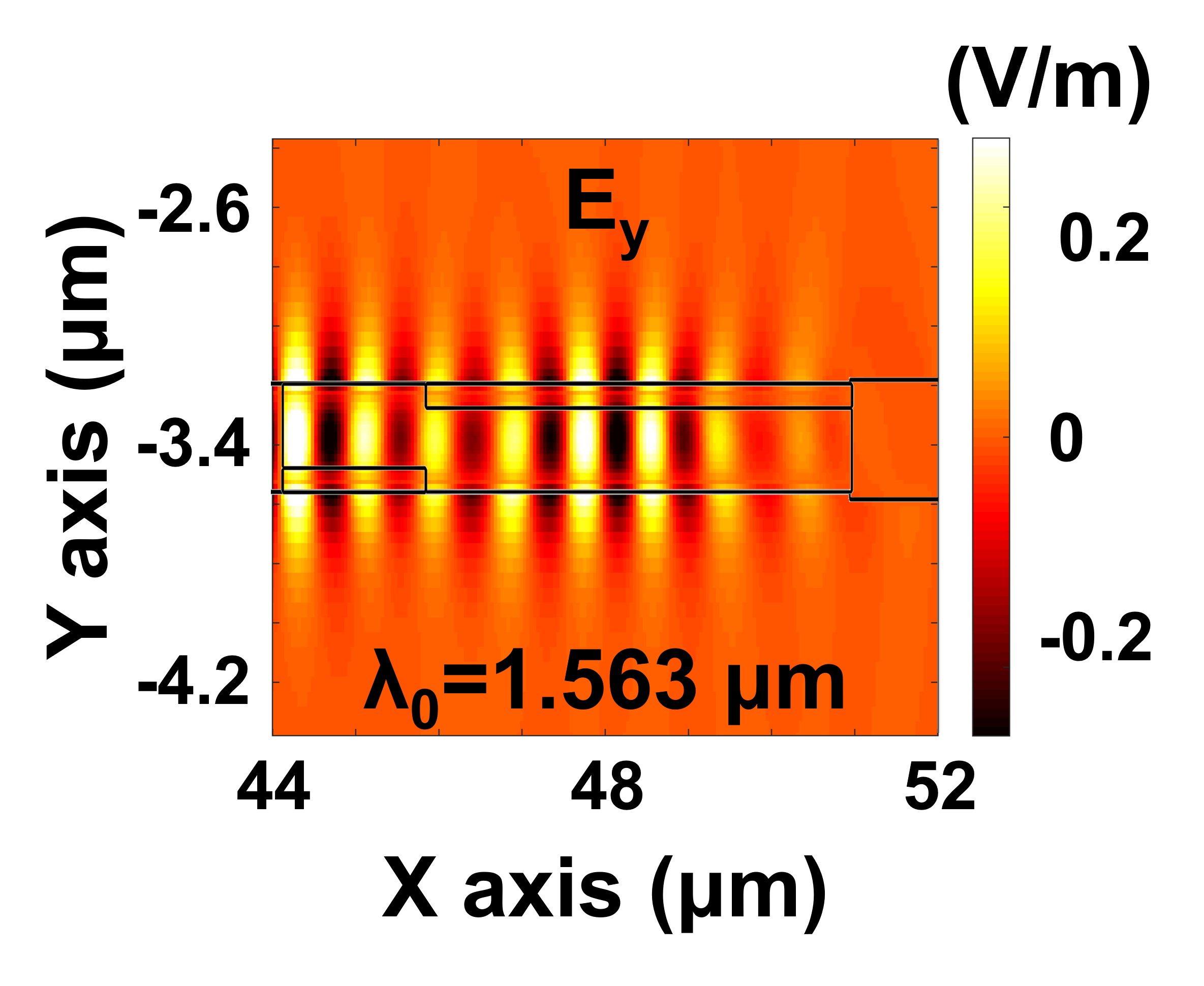}
                 \caption{}
                   \label{fig 4a}
                 \end{subfigure}
                 \begin{subfigure}[t!]{0.49\columnwidth}
                \centering
                 \includegraphics[scale=0.25,left]{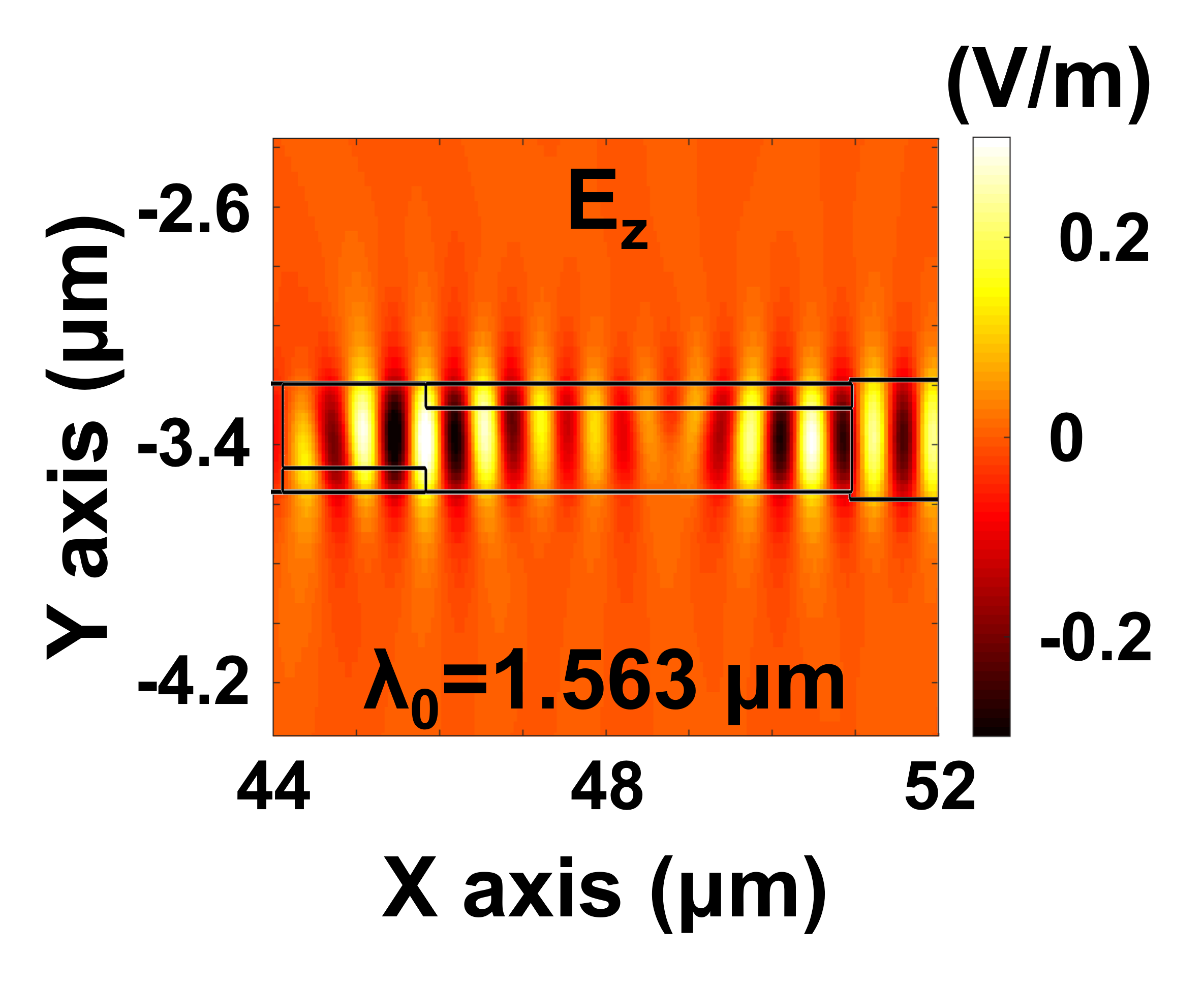}
                \caption{}
                 \label{fig 4b}
                \end{subfigure}
                 \begin{subfigure}[t!]{0.49\columnwidth}
                \centering
                \hspace*{0.01cm}
                 \includegraphics[scale=0.25,right]{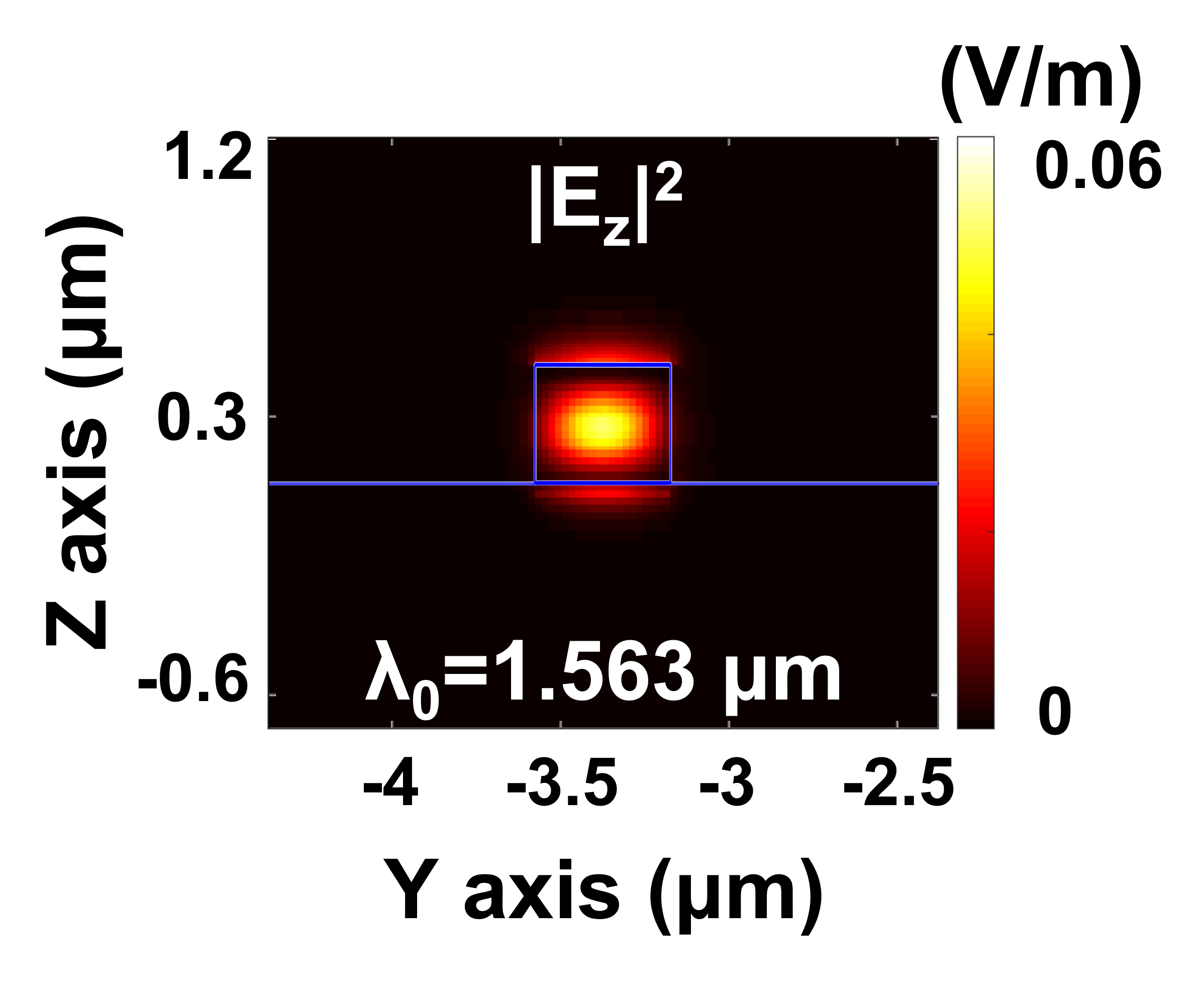}
                \caption{}
                 \label{fig 4c}
                \end{subfigure}
                 \begin{subfigure}[t!]{0.49\columnwidth}
                \centering
                \includegraphics[scale=0.25,left]{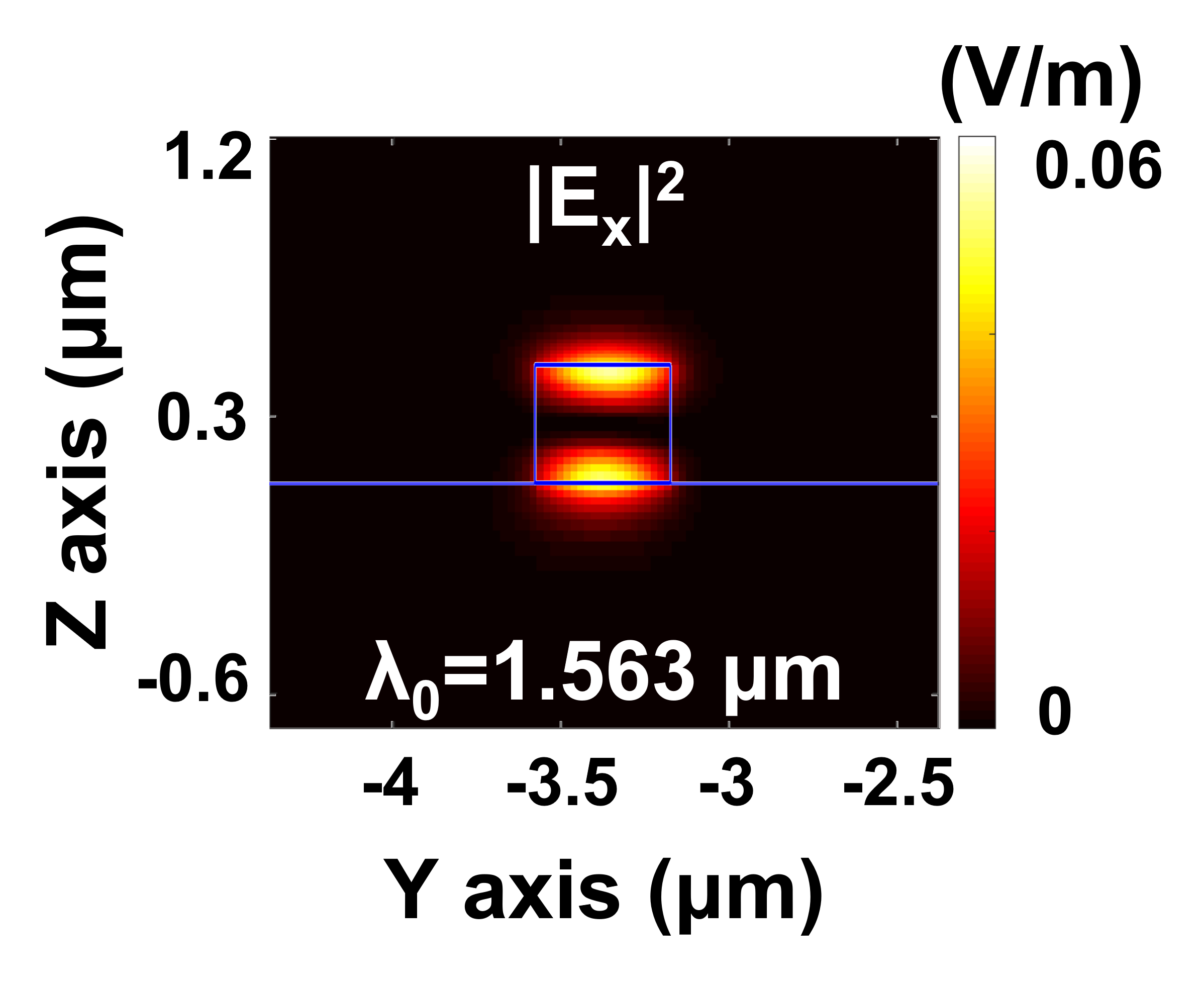}
                \caption{}
                 \label{fig 4d}
                \end{subfigure}
                \centering
                \caption{The 2D electric field distribution in the O4 waveguide for the two-section HWP at $\lambda_0$ = 1.563 $\mu$m. (a, b) The electric field components E\textsubscript{y} and E\textsubscript{z} of the light propagating through the waveguide, respectively. (c, d) The intensity profiles of the electric field components E\textsubscript{z} and E\textsubscript{x} respectively, illustrating the emergence of a TM\textsubscript{0} mode. The layout of the structures are also shown as an overlay.}
                \label{fig 4}
                \end{figure}
                
Figure \ref{fig 4} shows the performance of the two-section HWP by presenting the 2D electric field distribution through the O4 waveguide at $\lambda_0$ = 1.563 $\mu$m, where PCE is 99.95\%. Based on Figs. \ref{fig 4a} and \ref{fig 4b}, the distribution of the electric field components E\textsubscript{y} and E\textsubscript{z} along the propagation direction of light inside the waveguide illustrates the beating pattern between the two excited modes in the HWP region, indicated as an overlay in these figures (see Fig. \ref{fig 1c} for the layout of HWP). As elaborated in \cite{van2012increasing}, we designed our two-section HWP in a way that uses the combination of a QWP and a TQWP, it ultimately converts an input TE\textsubscript{0} mode to an output TM\textsubscript{0} mode with the highest possible PCE. Figures \ref{fig 4c} and \ref{fig 4d} show the intensity profiles of the electric field components E\textsubscript{z} and E\textsubscript{x} at the waveguide's output port of O4, where the latter is the longitudinal component of the waveguide mode. Both the profiles clearly show the existence of a TM\textsubscript{0} mode at the output.\par
                
\begin{figure}[h!]
             \centering
                \includegraphics[scale=0.3]{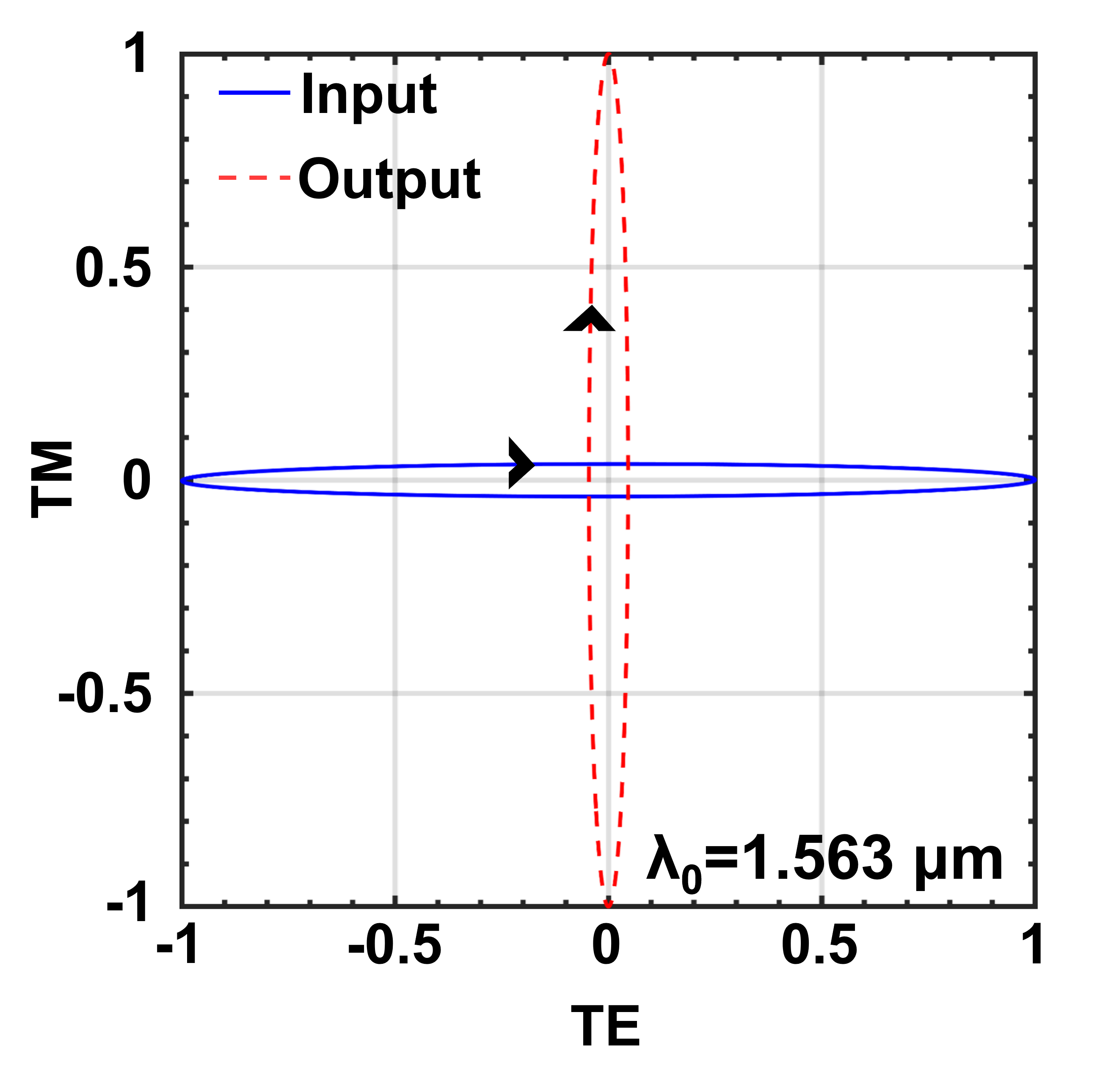}
                \centering
                \caption{Illustration of the polarization state of light using the polarization ellipse at $\lambda_0$ = 1.563 $\mu$m for the HWP.}
                \label{fig 5}
                \end{figure}

In order to quantitatively present the polarization state of the propagating light before and after the HWP, we used the polarization ellipse which provides information about both the polarization angle and ellipticity of the light mode. From the polarization ellipse of the input TE\textsubscript{0} mode in Fig. \ref{fig 5}, we obtained the polarization angle of $\theta$ = 0.07$^\circ$ and the ellipticity angle of $\chi$ = 2$^\circ$. But, at the output port of O4, the waveguide mode is rotated and the polarization angle reached to $\theta$ = 89.9$^\circ$ and $\chi$ = 4$^\circ$, resulting in the emergence of a TM\textsubscript{0} mode.\par

Based on the results presented in Figs. \ref{fig 4} and \ref{fig 5}, one can see the proper functionality of the proposed two-section HWP, while the results in Fig. \ref{fig 3} demonstrates that our two-section HWP outperforms the conventional one-section HWP.\par

\subsection{QWP Functionality} \label{ss-QWP}

             \begin{figure}[b!]
             \centering
               \includegraphics[scale=0.3]{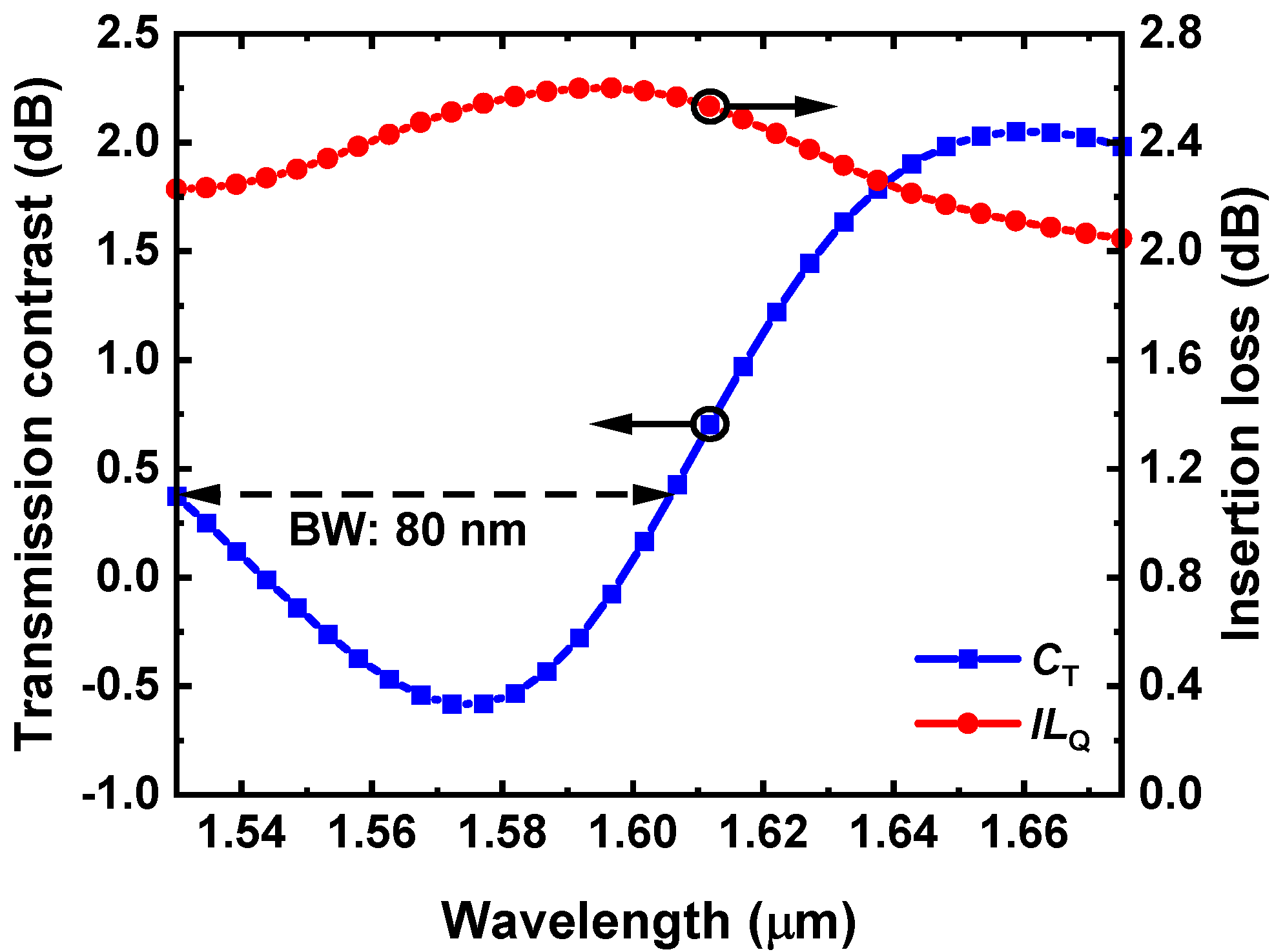}
                 \centering
                \caption{Performance of QWP over the telecom wavelength range of C to U, i.e. $\lambda_0$ = 1.53 to 1.675 $\mu$m. Transmission contrast and IL (\textit{C}\textsubscript{T} and \textit{IL}\textsubscript{Q}) at the left and right \textbf{Y} axes, respectively.}
                \label{fig 6}
                \end{figure}

The proposed device can also provide the QWP functionality as it delivers two quasi-circular polarized beams with opposite spins, by placing them mirrored relative to each other, one on the O2 waveguide and the other on the O3 waveguide. Because of this symmetry, we just show the performance of one of the QWP designs as shown in Fig. \ref{fig 6}. In order to have a QWP that properly works, the transmission contrast between the TM\textsubscript{0} and TE\textsubscript{0} modes must be $\approx$ 0 dB. According to the left \textbf{Y} axis of Fig. \ref{fig 6}, the transmission contrast of -0.6 dB $<$ \textit{C}\textsubscript{T} $< 0.6$ dB fulfills this criterion over the wavelength range of 1.53 to 1.61 $\mu$m, i.e. a bandwidth of 80 nm, which fully covers the C band and 55$\%$ of the C to U telecom bands. The right \textbf{Y} axis in this figure presents IL of the proposed QWP in which one can see that \textit{IL}\textsubscript{Q} varies between 2 to 2.6 dB across this ultrabroad wavelength range. It is noteworthy mentioning that the reflection is better than -20 dB over the same wavelength range. \par

\begin{figure}[b!]
             \centering
                \begin{subfigure}[t!]{0.49\columnwidth}
                \centering
                 \includegraphics[scale=0.25,right]{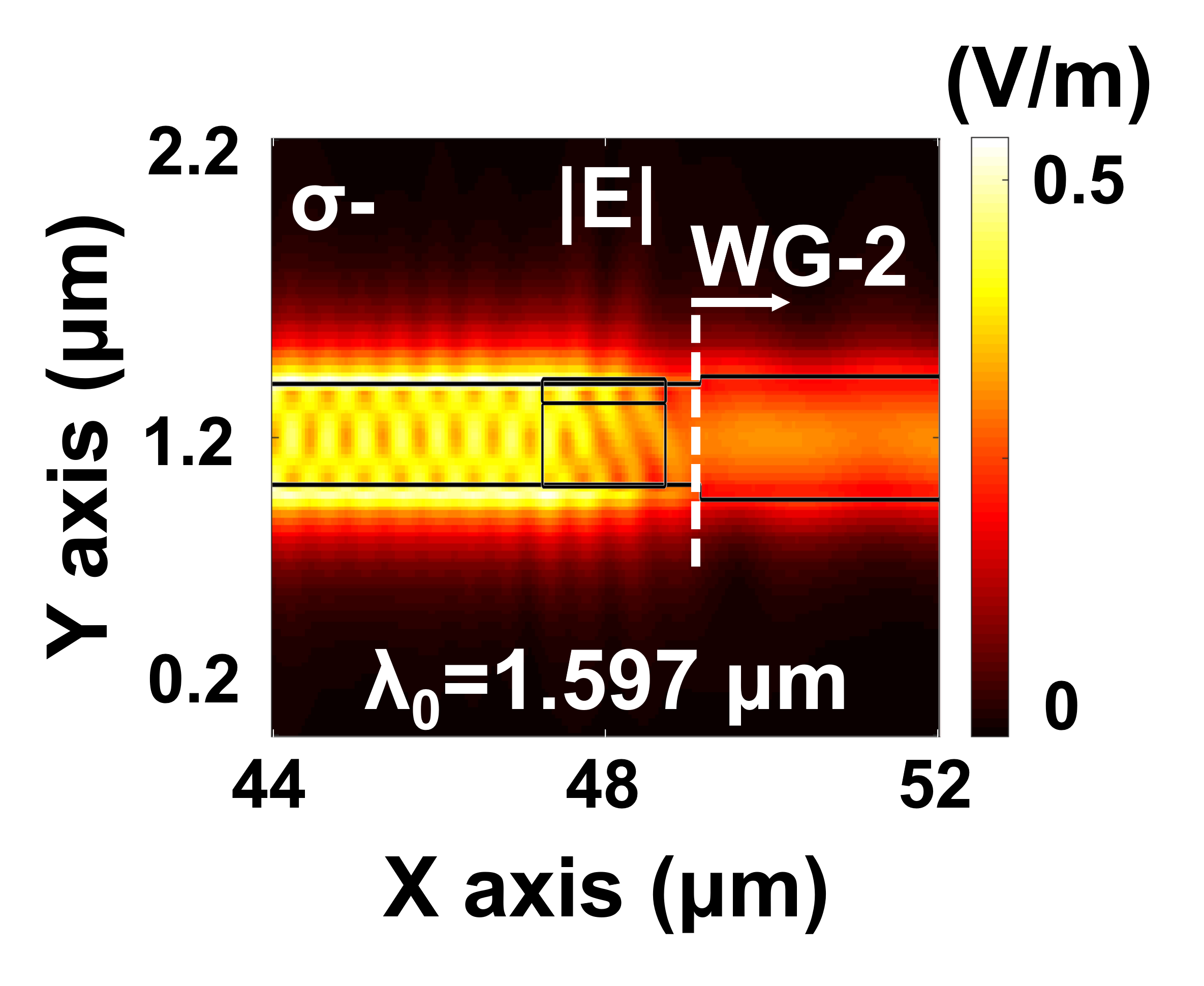}
                 \caption{}
                   \label{fig 7a}
                 \end{subfigure}
                 \begin{subfigure}[t!]{0.49\columnwidth}
                \centering
                 \includegraphics[scale=0.25,left]{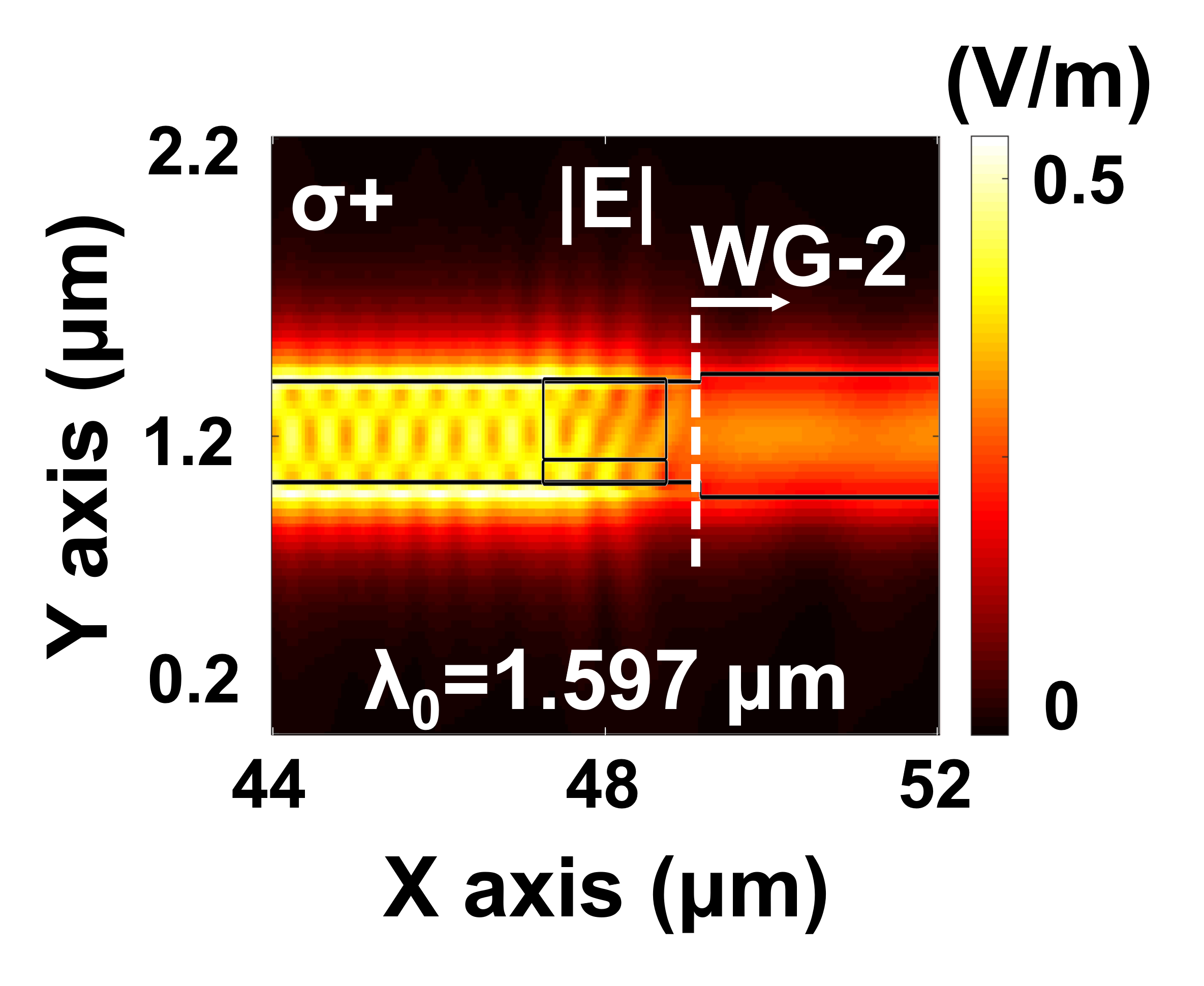}
                \caption{}
                 \label{fig 7b}
                \end{subfigure}
                 \begin{subfigure}[t!]{0.49\columnwidth}
                \centering
                 \includegraphics[scale=0.25,right]{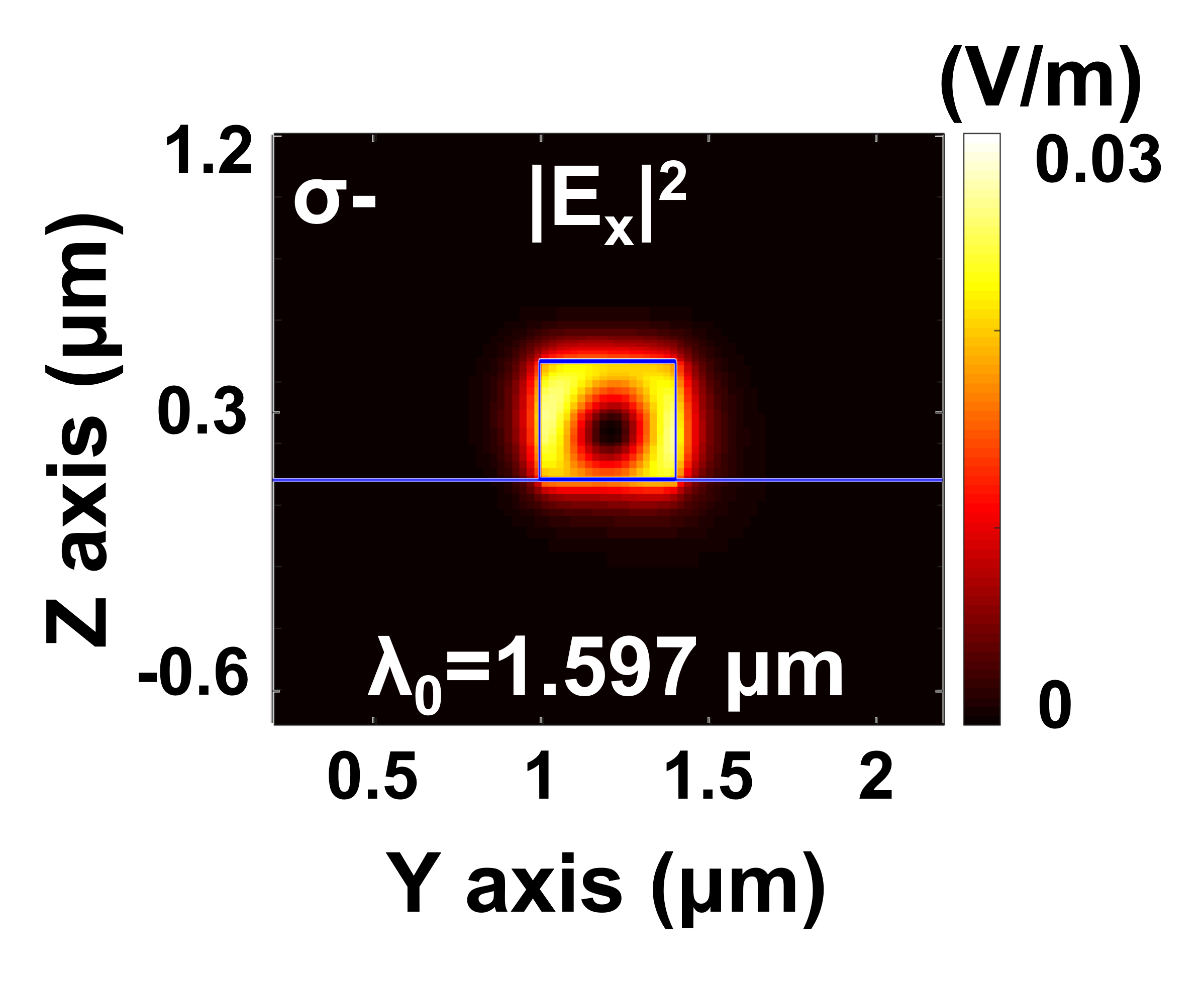}
                \caption{}
                 \label{fig 7c}
                \end{subfigure}
                 \begin{subfigure}[t!]{0.49\columnwidth}
                \centering
                 \includegraphics[scale=0.25,left]{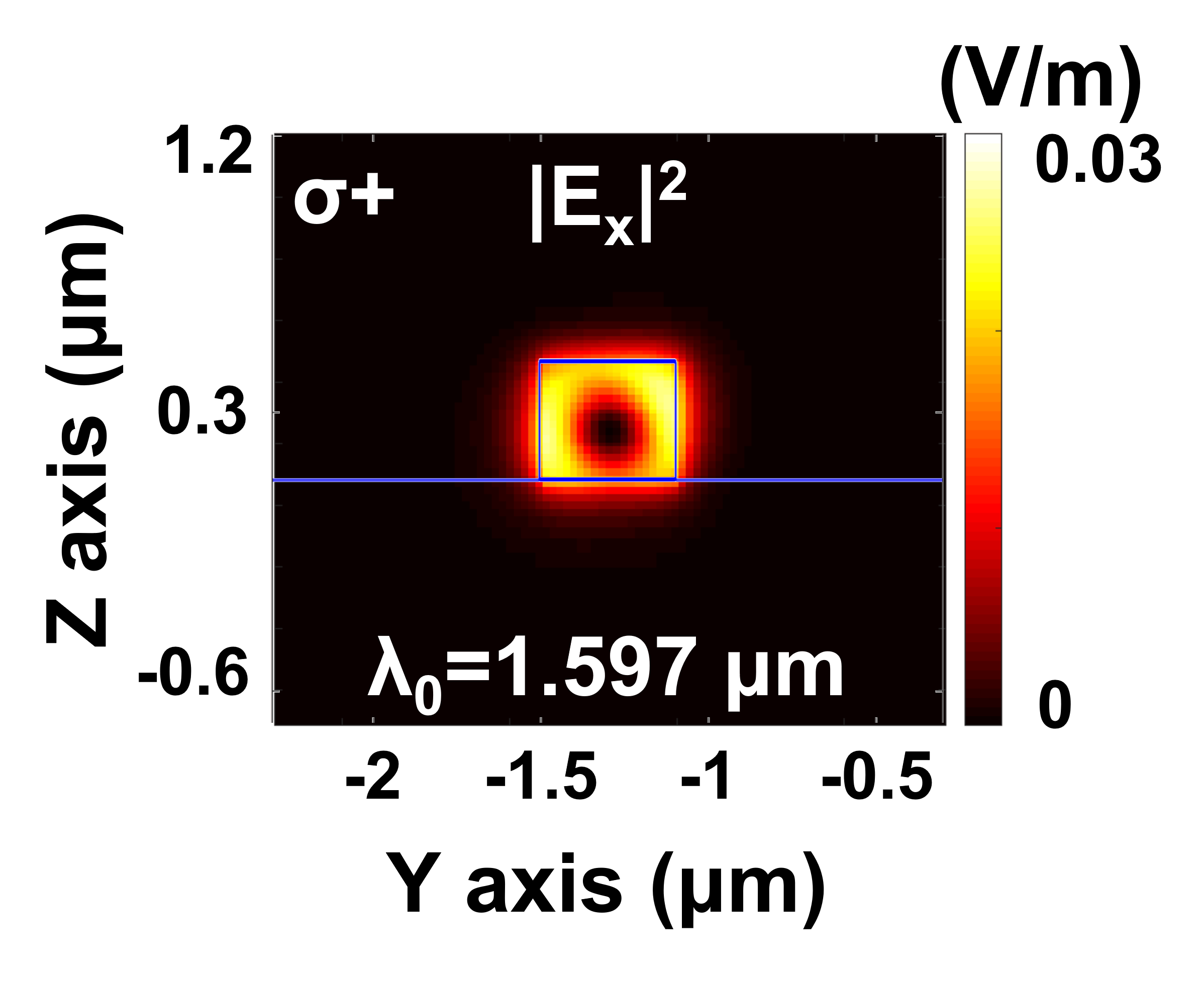}
                \caption{}
                 \label{fig 7d}
                \end{subfigure}
                \begin{subfigure}[t!]{0.49\columnwidth}
                \centering
                 \includegraphics[scale=0.25,right]{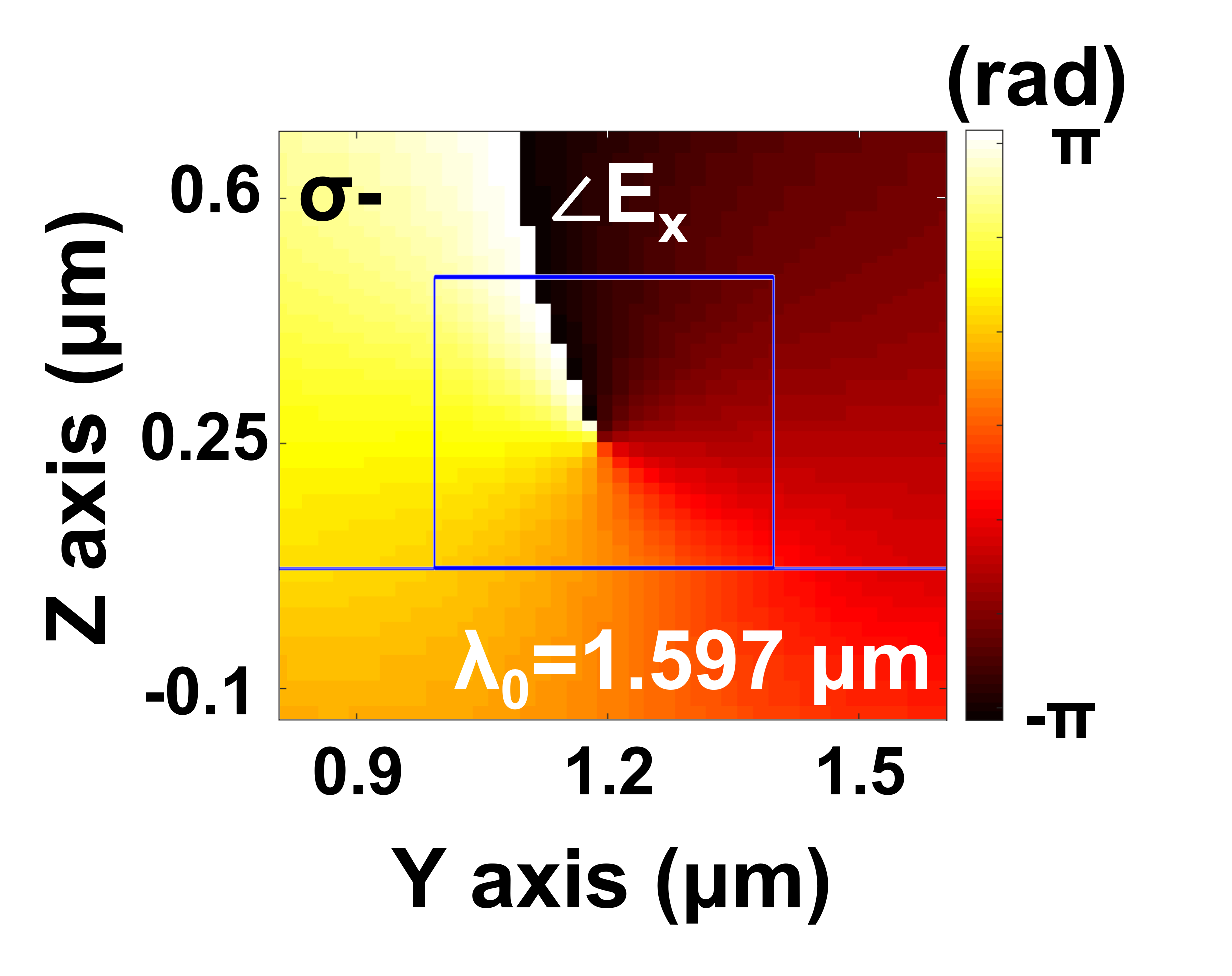}
                \caption{}
                 \label{fig 7e}
                \end{subfigure}
                \begin{subfigure}[t!]{0.49\columnwidth}
                \centering
                 \includegraphics[scale=0.25,left]{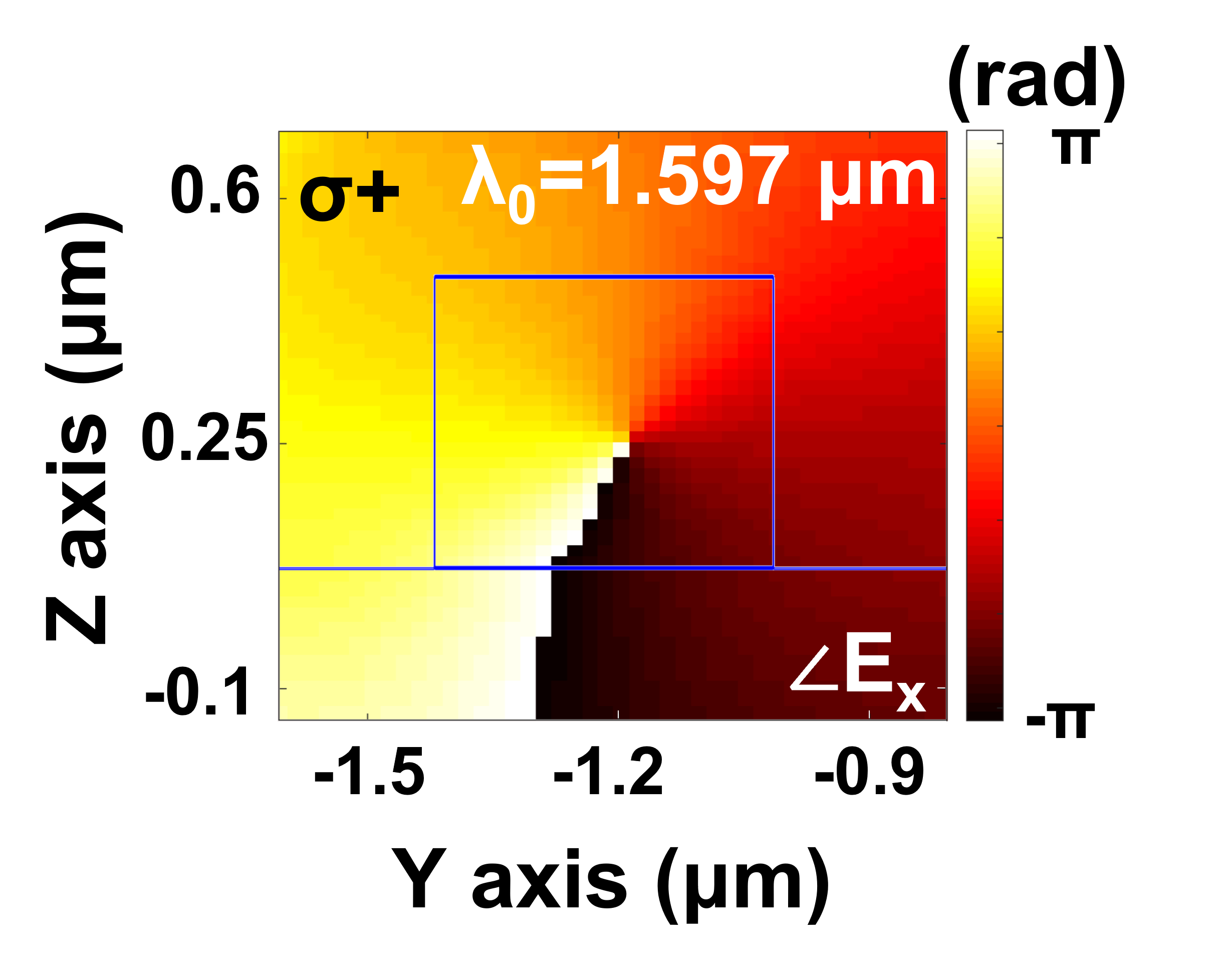}
                \caption{}
                 \label{fig 7f}
                \end{subfigure}
                \centering
                \caption{The 2D electric field distribution in the O2 and O3 waveguides for QWP at $\lambda_0$ = 1.597 $\mu$m. (a, b) The electric field magnitude $\lvert$E$\rvert$ of light propagating through the waveguide for two opposite spins of $\sigma-$ and $\sigma+$, respectively. (c-f) The intensity and phase profiles ($\angle$ E\textsubscript{x}) of the electric field component E\textsubscript{x} at the waveguide's output ports for $\sigma-$ and $\sigma+$, respectively. The plots in (c-f) altogether illustrate the longitudinal optical angular momentum of light with a topological charge of $l$ = $\pm$1, respectively. The layout of the structures are also shown as an overlay.}
                \label{fig 7}
                \end{figure}

The distributions of the electric field magnitude $\lvert$E$\rvert$ across the O2 and O3 waveguides are plotted in Figs. \ref{fig 7a} and \ref{fig 7b} at $\lambda_0$ = 1.597 $\mu$m, where \textit{C}\textsubscript{T} is 0 dB. As also denoted in these figures, it can be seen that the electric field spins in the plane of the propagation in both channels due to the transverse spin angular momentum (SAM) \cite{maltese2018towards}, where the light beam in O2 has a $\sigma-$ spin, while it has a $\sigma+$ spin in O3.\par
                
The field distributions in Figs. \ref{fig 7a} and \ref{fig 7b} also show that the profile of the propagating beam changes as it enters the second section of the waveguide (WG-2, see Fig. \ref{fig 1b}) as denoted by a white dashed line.  As mentioned in section \ref{s-Design}, we linked the O2 and O3 waveguides with two waveguides with different widths, where the second part with the width of $d$\textsubscript{WG-2} is designed to suppress the birefringence in the waveguide effectively. In this way, we will have quasi-circular polarized light beams in O2 and O3 that carry longitudinal orbital angular momentum (OAM) \cite{maltese2018towards,ni2019selective}. To illustrate this more, we presented the intensity and phase profiles of the longitudinal component of the electric field, i.e. E\textsubscript{x}, in Figs. \ref{fig 7c} to \ref{fig 7f} for both O2 and O3, respectively. According to Figs. \ref{fig 7c} and \ref{fig 7d}, there is a point of zero intensity in the intensity profiles of E\textsubscript{x} in both outputs. In addition, the phase profiles of E\textsubscript{x} in O2 and O3 have a spiral-shaped front as presented in Figs. \ref{fig 7e} and \ref{fig 7f}. Both these features are characteristics of longitudinal OAM in a photonic waveguide \cite{liang2016integratable,maltese2018towards,ni2019selective}. It is worth mentioning that the detailed explanation on SAM and OAM is out of the scope of this work and interested readers are referred to \cite{aiello2015transverse,bliokh2015spin,bliokh2015transverse} for detailed description. Overall, one can see that our PC device can provide another degree of freedom by creating two optical beams carrying SAM and OAM with $l$ = $\pm$1 from only a TE\textsubscript{0} polarized light beam without changing the propagation direction.\par

\begin{figure}[ht!]
             \centering
                \begin{subfigure}[t!]{0.48\columnwidth}
                 \includegraphics[scale=0.235,right]{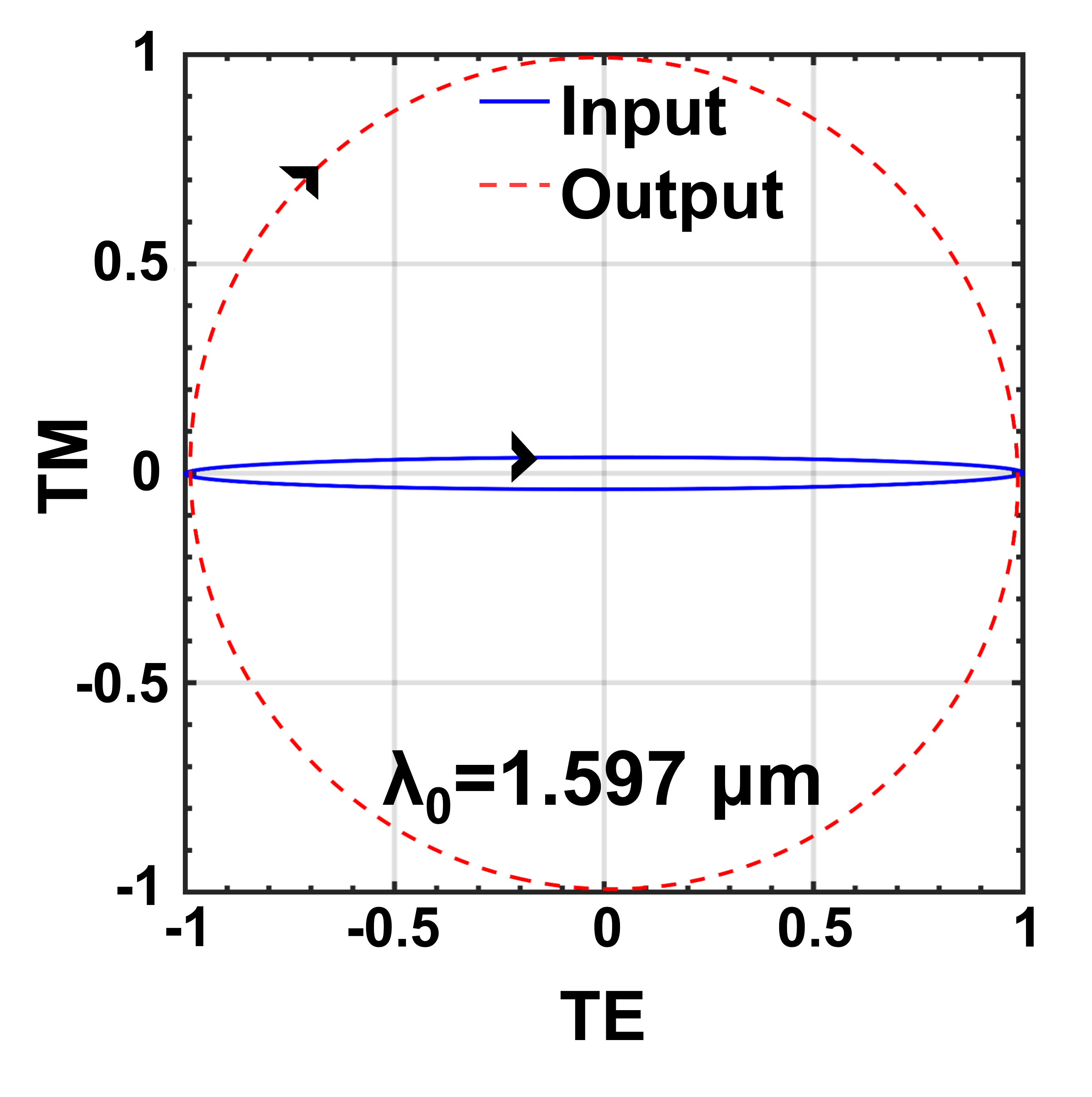}
                 \caption{}
                   \label{fig 8a}
                 \end{subfigure}
                 \begin{subfigure}[t!]{0.48\columnwidth}
                \includegraphics[scale=0.235,left]{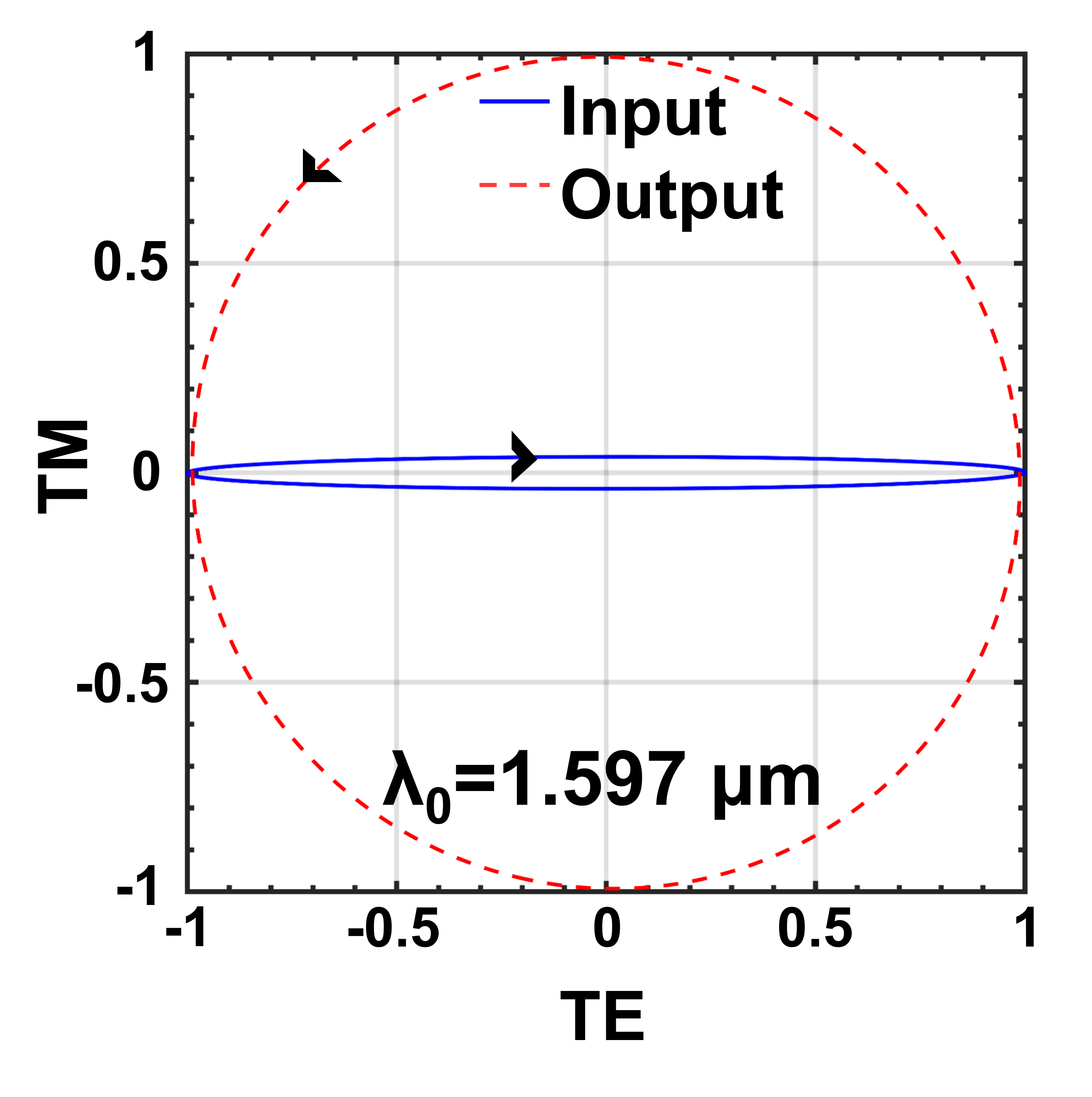}
                \caption{\hspace*{-1cm}}
                 \label{fig 8b}
                \end{subfigure}
                \centering
                \caption{Illustration of the polarization state of the light using polarization ellipse at $\lambda_0$ = 1.597 $\mu$m for the QWP. (a, b) $\sigma-$ and $\sigma+$ spins in the O2 and O3 waveguides, respectively.}
                \label{fig 8}
                \end{figure}

Similar to what is done in section \ref{ss-HWP}, in order to quantitatively illustrate the change in the polarization state of the input light, the polarization ellipse is used, for which the results are presented in Fig. \ref{fig 8}. Likewise, we have a TE\textsubscript{0} input light mode which is converted to two quasi-circular polarized beams. Based on the results, the ellipticity and polarization angles of the output fields of the O2 and O3 waveguides are equal to $\chi$ = 45$^\circ$ and $\theta$ = 45$^\circ$.\par

\section{Discussion} \label{s-Discussion}

Based on the results in section \ref{s-results}, we illustrated that our device can offer a HWP functionality. This device converts an input TE\textsubscript{0} mode into a TM\textsubscript{0} mode with an efficiency of $\geq$ 91\% over the entire C to U telecom bands, while it offers a 0.2 dB bandwidth of 125 nm with a PCE of $\geq$ 95\%. Note that a PCE of $\geq$ 95\% corresponds to a polarization extinction ratio of larger than 13 dB. In comparison to the recently published works on HWPs \cite{zhang2011ultracompact,caspers2012compact,komatsu2012compact}, our device benefits from very efficient conversion efficiency, very large polarization extinction ratio, and ultra broadband operational window. Moreover, due to the application of the two-section technique, our HWP can also offer high tolerance to fabrication errors as demonstrated in \cite{van2012increasing}, while keeping IL below 8 dB. It is noteworthy mentioning that the IMOS-platform offers semiconductor optical amplifier (SOA)-functionality to compensate the losses.\par

In addition, our proposed device can efficiently create two quasi-circular polarized light beams with opposite spins. Results showed that this device can maintain QWP functionality over the whole C band and 55\% of the C to U telecom bands, while IL remains below 2.6 dB. In contrast, in the recent reports on integrated QWPs \cite{zhang2011ultracompact,gao2015chip,guo2016pure,liang2016integratable,maltese2018towards,ni2019selective}, broadband operation of QWP, and the possibility of having light beams with two opposite spins on the same chip without reversing the propagation direction of a TE\textsubscript{0} or a TM\textsubscript{0} light mode have not been investigated. From the fabrication perspective, if common fabrication errors give rise to birefringence in the output waveguides O2 and O3, we can employ phase-change materials, e.g. germanium-antimony-tellurium (GST), to control and minimize the birefringence in the waveguides \cite{rude2013optical}. \par

The proposed device in this paper illustrates polarization demultiplexing with a wide applications range from telecom to biosensing. From an input TE\textsubscript{0} mode, our device can create a TM\textsubscript{0} light mode and two quasi-circular polarized light beams. The TE\textsubscript{0} to TM\textsubscript{0} conversion can be served as a polarization degree of freedom for quantum information processing. The generated light beams by QWPs can carry SAM and OAM with $l$ = $\pm$1, which can be quite useful in quantum information processing by offering the OAM degree of freedom, i.e. encoding signals with opposite topological charges. \par

\section{Conclusion}
To conclude, we numerically demonstrated a multifunctional integrated plasmonic-photonic device based on a compact 1$\times$4 MMI for polarization demultiplexing in the IMOS platform. This device can offer the HWP and QWP functionalities with high efficiency over the 86.2\% and 55\% of the C to U telecom bands, i.e. $\lambda_0$ = 1.53 to 1.675 $\mu$m, respectively. Our device with such characteristics can be a potential building block for controlling the light polarization in PICs, which not only is important for on-chip telecom applications, but also can have advantages for biosensing area. Finally, we expect that the device in this paper can play as a key photonic component for quantum information processing by offering both polarization and angular momentum degrees of freedom.\par


%

\ifCLASSOPTIONcaptionsoff
  \newpage
\fi



%
\bibliographystyle{IEEEtran}
\bibliography{MyLib}

%

\vskip -2\baselineskip plus -1fil
%

\begin{IEEEbiography}[{\includegraphics[width=1in,height=1.25in,clip,keepaspectratio]{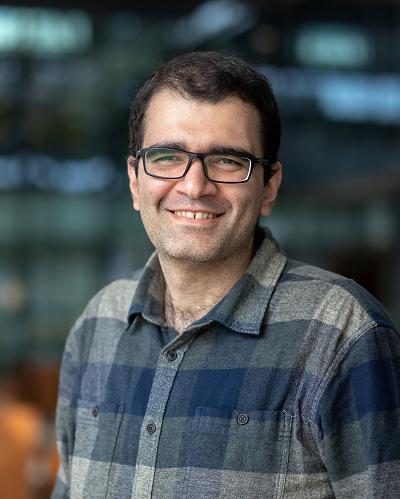}}]{Hamed Pezeshki}
received his M.Sc from IAU, Science and Research Branch, Tehran, Iran in Feb. 2012. His research was in the field of photonic crystals with applications in telecommunications, for which he received a financial grant from the Iran Nanotechnology Initiative Council. Then, in Dec. 2016, he joined the Optics and Photonics group in the Department of Electrical and Electronic engineering at the University of Nottingham, Nottingham, United Kingdom, where he received the "Faculty of Engineering Research Excellence PhD Scholarship" for his course of study. His research was focused in the area of plasmonics with applications in biophotonics and telecommunications. Currently, he is a post-doctoral researcher in the Department of Applied Physics, Physics of Nanostructure group at the Eindhoven University of Technology. His current research is about the development of hybrid spintronic-photonic memories in collaboration with the Institute of Photonic Integration at the Eindhoven University of Technology. His research interests include photonic crystals, plasmonic nanostructures, and epsilon-near-zero materials.
\end{IEEEbiography}

\vskip -2\baselineskip plus -1fil

\begin{IEEEbiography}[{\includegraphics[width=1in,height=1.25in,clip,keepaspectratio]{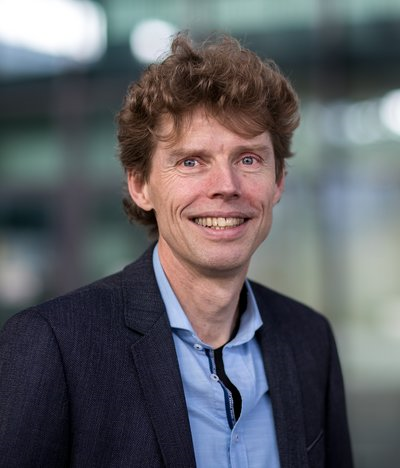}}]{Bert Koopmans}
(1963) graduated and received his PhD degree at the University of Groningen. After a short stay as a postdoc at the Radboud University Nijmegen, he spent three years as a Humboldt Fellow at the Max-Planck Institute for Solid State Physics in Stuttgart. In 1997 he joined the Eindhoven University of Technology, where since 2003 he chairs the Group Physics of Nanostructures. His current research interests encompass spintronics, nanomagnetism and ultrafast magnetization dynamics. He participates in the NWO Gravitation program on integrated nanophotonics, initiating research on hybrid spintronic-photonic devices. In 2020 he was elected Distinguished Lecturer of the IEEE Magnetics Society.
\end{IEEEbiography}

\vskip -2\baselineskip plus -1fil

\begin{IEEEbiography}[{\includegraphics[width=1in,height=1.25in,clip,keepaspectratio]{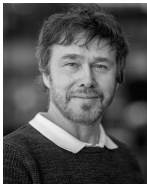}}]{Jos J. G. M. van der Tol}
received the M.Sc. and Ph.D. degrees in physics from the State University of Leiden, Leiden, The Netherlands, in 1979 and 1985, respectively. In 1985, he joined KPN Research, where he had been involved in the research on integrated optical components for use in telecommunication networks. Since July 1999, he has been an Associate Professor with the Eindhoven University of Technology, Eindhoven, The Netherlands, where his research interests include opto-electronic integration, polarization issues, photonic membranes, and photonic crystals. He has coauthored more than 250 publications in the fields of integrated optics and optical networks and has 25 patent applications to his name. He has been working on guided wave components on III–V semiconductor materials. He has also been active in the field of optical networks, focusing on survivability, introduction scenarios, and management issues. His research interests include modeling of waveguides, design of electro-optical devices on lithium niobate, and their fabrication.
\end{IEEEbiography}




\end{document}